\documentclass[12pt]{article}
\usepackage[table]{xcolor}
\usepackage{amsmath,amssymb,exscale,cancel}
\usepackage{slashed}
\usepackage{graphicx}
\usepackage{epsfig}
\usepackage{multicol}
\usepackage{color}
\usepackage{mathrsfs}
\usepackage{blindtext}
 \usepackage{fancyhdr}
\usepackage{hyperref}
\usepackage{cite}
\usepackage{mathtools}

\usepackage{floatrow}

\usepackage{graphicx}
\usepackage{sidecap}

\usepackage[latin1]{inputenc} 

\usepackage{multicol}  
 
 \usepackage{titlesec}
 
 \usepackage{rotating,slashed,xcolor,amsfonts,expdlist,charter}

\numberwithin{equation}{section}

\usepackage{xcolor}
\usepackage{sectsty}

\usepackage{mdframed}
\usepackage{titletoc}

\numberwithin{equation}{section}

\usepackage{xcolor}
\usepackage{sectsty}

\usepackage{mdframed}
\usepackage{titletoc}

\definecolor{secnum}{RGB}{13,151,225}
\definecolor{ptcbackground}{RGB}{212,237,252}
\definecolor{ptctitle}{RGB}{0,177,235}

\titlecontents{lsection}
  [5.8em]{\sffamily}
  {\color{secnum}\contentslabel{2.3em}\normalcolor}{}
  {\titlerule*[1000pc]{.}\contentspage\\\hspace*{-5.8em}\vspace*{5pt}%
    \color{white}\rule{\dimexpr\textwidth-15.5pt\relax}{1pt}}

\usepackage{hyperref}
\hypersetup{colorlinks,bookmarksopen,bookmarksnumbered,citecolor=rossos,
linkcolor=redy,pdfstartview=FitH,urlcolor=green-go}
\usepackage{slashed}

\definecolor{blus}{cmyk}{1,0.9,0,0.1}
\definecolor{verdes}{cmyk}{0.99,0,0.59,0.65}
\definecolor{rossos}{cmyk}{0,1,1,0.55}
\definecolor{redy}{cmyk}{0,1,1,0.7}
\definecolor{greeny}{cmyk}{0.99,0,0.59,0.98}
\definecolor{green-go}{cmyk}{0.79,0,0.59,0.5}

\usepackage{titlesec}

\newcommand{\beq}{\begin{equation}}
\newcommand{\eeq}{\end{equation}}
\newcommand{\3}{{\hspace{-0.4cm}3\hspace{0.25cm}}}

\def\hhref#1{\href{http://arxiv.org/abs/#1}{arXiv:#1}} % in bibliography

 \def\Lag{\mathscr{L}}
 
\newcommand{\tmtextbf}[1]{{\bfseries{#1}}}
\newcommand{\tmtextrm}[1]{{\rmfamily{#1}}}
\newcommand{\bp}{\bar M_P}

\def\be{\begin{equation}}
\def\ee{\end{equation}}
\def\ba{\begin{array} }

\def\bac{\begin{array} {c}}
\def\bacc{\begin{array} {cc}}
\def\baccc{\begin{array} {ccc}}
\def\bacccc{\begin{array} {cccc}}
\def\ea{\end{array}}
\def\bea{\begin{eqnarray}}
\def\eea{\end{eqnarray}}

\definecolor{red}{rgb}{1,0,0}

\def\psl{\hbox{\hbox{${p}$}}\kern-1.9mm{\hbox{${/}$}}}
\def\dsl{\hbox{\hbox{${\partial}$}}\kern-2.2mm{\hbox{${/}$}}}
\def\Dsl{\hbox{\hbox{${D}$}}\kern-2.6mm{\hbox{${/}$}}}

\newcommand{\gappeq}{{\rlap{{\raise}.5ex\text{\ensuremath{>}}}{{\lower}.5ex\text{\ensuremath{\sim}}}}}
\newcommand{\lappeq}{{\rlap{{\raise}.5ex\text{\ensuremath{<}}}{{\lower}.5ex\text{\ensuremath{\sim}}}}}
\newcommand{\I}{\tmtextrm{1{\kern}-.24em l}}

\begin{document}
\topmargin -1.0cm
\oddsidemargin 0.9cm
\evensidemargin -0.5cm

{\vspace{-1cm}}
\begin{center}

\vspace{-1cm}

%TO DO
%introduction x
%ordinare citazioni x
%controllare coerenza simboli x
%add refs for JCAP edit. x
%conclusions x
%abstract x
%add refs using 
%  - iNSPIRE (a salvio) x
%  - the literature folders of "Hamiltonian formulation and canonical form" x
%  - that of "no-scale gravity" x
%check logic and formulae of the paper part x
%acronimi x

%- split into notes and paper,  add acknowledgements and eliminate references not cited in the paper x

%- check English of paper x

%- print and read paper x

%- upload on arXiv x

 {\LARGE \tmtextbf{ 
\color{verdes} \hspace{-0.9cm}    \\
A non-Perturbative  and Background-Independent Formulation of Quadratic Gravity \hspace{-1.6cm}}} {\vspace{.5cm}}\\

\vspace{1.3cm}

{\large  {\bf Alberto Salvio }
{\em  
\vspace{.4cm}

 Physics Department, University of Rome Tor Vergata, \\ 
via della Ricerca Scientifica, I-00133 Rome, Italy\\

\vspace{0.6cm}

I. N. F. N. -  Rome Tor Vergata,\\
via della Ricerca Scientifica, I-00133 Rome, Italy\\

\vspace{.4cm}

%\today 

\vspace{0.4cm}

\vspace{0.2cm}

 \vspace{0.5cm}
}

\vspace{.3cm}

}
\vspace{0.cm}

\end{center}

%
% \begin{\large abstract}
% 
\noindent ---------------------------------------------------------------------------------------------------------------------------------
\begin{center}
{\bf \large \color{verdes}Abstract}
\end{center}
\noindent A non-perturbative and background-independent quantum formulation of quadratic gravity is provided. A canonical quantization procedure introduced in previous works, named after Dirac and Pauli, is here applied to quadratic gravity to obtain, as required by consistency, a well-defined Euclidean path integral. The theory is unitary: all probabilities are non negative and they sum up to one. We obtain path-integral expressions for the transition amplitudes, Green's functions and generic matrix elements of time-ordered products of the metric. As a byproduct, similar results are also obtained for a scalar-field four-derivative interacting model. In this way, among other things, previous perturbative and background-dependent calculations are  justified.   The (quantum) quadratic gravity effective action, whose field equations determine the vacuum expectation value of the metric in the presence of a generic energy-momentum tensor, is constructed. The classical limit of the effective action turns out to be equivalent to the starting classical action of quadratic gravity, whose runaway rates were previously shown to be slow enough to be compatible with observations. Finally, the constructed non-perturbative and background-independent quantum quadratic gravity is applied to quantum cosmology to obtain a path-integral expression for the wave function of the universe, which satisfies a sort of Wheeler-DeWitt equation. This application allows us to understand at the quantum level why our universe is nearly homogeneous and isotropic.

\vspace{0.7cm}

\noindent---------------------------------------------------------------------------------------------------------------------------------

  \vspace{-0.9cm}
  
  \newpage 
\tableofcontents

\noindent

\vspace{0.5cm}

%\section{Introduction}\label{Introduction}

\section{\color{verdes} Introduction}\label{intro}

Einstein's general relativity (GR) is an extremely successful classical theory of gravity that has been tested by several observations: recent tests include the  discovery of gravitational waves, whose production is consistent with coalescing black holes~\cite{LIGOScientific:2016aoc}, and the images of the black holes in the center of  our galaxy and the M87 one produced by the Event Horizon Telescope~\cite{EventHorizonTelescope:2019dse,EventHorizonTelescope:2019ggy,EventHorizonTelescope:2022xnr,EventHorizonTelescope:2022wok,EventHorizonTelescope:2022xqj}. Furthermore, GR, appropriately extended by matter fields, is at the basis of our current understanding of cosmology, including the early universe.

However, an extension of GR is necessary because of quantum mechanics: GR is known to be nonrenormalizable by perturbative methods~\cite{Goroff:1985sz,Goroff:1985th} and to be within the regime of validity of perturbation theory at length scales much below the Planck one. Thus, while it is possible that GR might make sense as a quantum theory at Planckian and trans-Planckian scales through some (yet unknown) non-perturbative approach, one must still extend it to make contact with large-distance physics. 

Another shortcoming of quantum GR is the lack of a consistent Euclidean path integral: the Euclidean action of GR is unbounded from below, an issue that goes under the name of the ``conformal-factor problem". This is a serious issue as all known quantum field theories of physical relevance are consistently defined only  as analytic continuations of the corresponding Euclidean theories.

Both the non-renormalizability and the conformal-factor problem of GR could be solved by adding quadratic-in-curvature terms. Indeed, the resulting theory, which is often called quadratic gravity\footnote{Other used names include higher-derivative gravity and four-derivative gravity. See~\cite{Salvio:2018crh,Donoghue:2021cza} for reviews.}, is renormalizable~\cite{Weinberg:1974tw,Deser:1975nv,Stelle:1976gc,Barvinsky:2017zlx} and a simple application of the prescription~\cite{Gibbons:1994cg} to determine the Euclidean action indicates that is also free from the conformal-factor problem in a sizable region of its parameter space~\cite{Menotti:1989ms}. 

A non-perturbative quantization leading to a well-defined Euclidean path integral has been found in Ref.~\cite{Salvio:2015gsi} in the context of a simple, yet non-trivial, toy model for quadratic gravity: an interacting version of the Pais-Uhlenbeck model~\cite{Pais}. This quantization is based on the quantum treatment of canonical variables first discussed by Pauli~\cite{Pauli}, who elaborated on a previous work by Dirac~\cite{Dirac}; for this reason it was called the Dirac-Pauli (DP) quantization in~\cite{Salvio:2015gsi}. The model discussed in~\cite{Salvio:2015gsi}, however, features a finite number of degrees of freedom, so a  question that has remained unanswered so far is whether the DP quantization can be extended to field theory, which features instead an infinite number of degrees of freedom, and ultimately to quadratic gravity. 
In Refs.~\cite{Buchbinder-Lyahovich:1987,Buchbinder:1991ne,Buchbinder:1992rb} a real-time path integral was constructed for quadratic gravity using an ordinary (non-DP) quantization. As we will show here, this ordinary quantization does not allow us to construct a consistent Euclidean path integral.

As discussed in~\cite{Salvio:2015gsi,Salvio:2018crh}, the DP quantization implies the presence of an indefinite inner product. So one must ask whether consistent probabilities (non-negative numbers that sum up to one) can be defined in such a situation; in other words, one must ask whether  this quantization respects unitarity. The answer is yes, as shown in Refs.~\cite{Strumia:2017dvt,Salvio:2019wcp,Salvio:2020axm}.  Indeed, by applying the standard frequency definition of probability as in real physics experiments, it turns out that in the Born rule one should not use the indefinite  inner product but instead positive-definite inner products, which can be constructed by inserting appropriate linear operators inside the indefinite inner product. We will refer to these operators as norm operators. 

The existence of a different inner product in relation with the unitarity of the theory has been discussed in a number of papers in the context of quantum mechanics and field theory with an antilinear symmetry of the Hamiltonian (see e.g.~Refs.~\cite{Bender:2008gh,Bender:2007wu} for applications to the Pais-Uhlenbeck oscillator and Ref.~\cite{Bender:2023cem} for a review). 

Norm operators have been used so far to provide a unitary quantization of quadratic gravity through a perturbative and background dependent approach: one starts from a fixed classical background (e.g.~the Minkowski or the Friedmann-Lemaitre-Robertson-Walker (FLRW) metrics) and then considers a semiclassical expansion where the perturbations are quantized with these rules~\cite{Salvio:2017xul,Salvio:2019wcp, Salvio:2020axm}. A perturbative and background-dependent approach is present in several roads to quantum gravity, including string theory. Needless to say, such an approach must ultimately be replaced by a non-perturbative and background-independent formulation. Indeed, one of the most important Einstein's messages on gravity is that the laws of physics should be formulated in a background-independent way. The purpose of this paper is to find a non-perturbative formulation of this type in quadratic gravity.

Note that in the perturbative background-dependent approach used so far, one first starts from the classical theory. As will be discussed in detail in this work, the classical Hamiltonian of quadratic gravity is unbounded from below (and above). In this sense quadratic gravity belongs to the class of higher-derivative models studied by Ostrogradsky in Ref.~\cite{ostro}.   Nevertheless, in~\cite{Salvio:2019ewf,Gross:2020tph} (see also~\cite{Held:2021pht,Held:2023aap} for  related works) it has been found that the runaway rates in classical quadratic gravity can be slow enough\footnote{In some simple cases it has even been possible to show that Ostrogradsky's unboundedness is compatible with absolute stability~\cite{Deffayet:2021nnt,Deffayet:2023wdg}.} to ensure agreement with observations. This occurs when the typical length scales are much above a threshold that is low enough to describe the whole cosmology, i.e.~the resulting theory can be called metastable rather than unstable, see Ref.~\cite{Salvio:2019ewf}. The possible instability that takes place when such bound is violated not only is compatible with cosmology, but would also explain why we live in a homogeneous and isotropic universe~\cite{Salvio:2019ewf}: this is because this instability can be effectively activated only in sufficiently inhomogeneous and anisotropic patches of the universe. 

The non-perturbative and background independent formulation that we seek in this paper can, among other things, provide us with a fully quantum and non-perturbative justification of these classical cosmological findings. Furthermore, such a formulation can also be used to make contact with other (apparently independent) perturbative and background-dependent approaches to formulate quadratic gravity as a unitarity and (meta)stable renormalizable theory: these approaches include the Lee-Wick~\cite{Lee:1969fy,Salvio:2018kwh,Donoghue:2019fcb,Holdom:2021hlo} and the fakeon~\cite{Anselmi:2017ygm} programs, where all observable degrees of freedom are quantized  with standard rules.

 Even in ordinary quantum theories, the real-time path integral is only a formal object, whose consistency at the rigorous level is questionable. Therefore, we seek here a non-perturbative and background-independent formulation of quadratic gravity with a consistent Euclidean path integral. The real-time (Lorentzian) physics should then be {\it defined}, like in all known physically relevant quantum field theories, as an analytic continuation of the Euclidean  theory. A formulation of this type can then be used to give a rigorous basis for quantum cosmology, where, as proposed by Hartle and Hawking, one describes everything through a quantum mechanical ``wave function of the universe"~\cite{Hartle:1983ai}. Indeed, the Euclidean path integral can  be used to obtain eigenfunctions of the Hamiltonian that solve the Shroedinger equation. In the context of quantum cosmology this equation is known as the Wheeler-DeWitt equation~\cite{DeWitt:1967yk,Wheeler}. A consistent quantum cosmology in quadratic gravity would be interesting  also because this theory contains Starobinsky's model of inflation~\cite{Starobinsky:1980te}, which is currently in perfect agreement with the cosmic-microwave-background observations~\cite{Ade:2015lrj,Planck2018:inflation,BICEP:2021xfz}.
 
 Yet another application of a non-perturbative and background-independent formulation of quadratic gravity would be in the context of asymptotic safety~\cite{WeinbergAS}, where one attempts to render gravity UV complete through a non-perturbative UV fixed point. Fixed points of this type have been found for quadratic gravity in Refs.~\cite{Benedetti:2009rx,Falls:2020qhj}.
 
\vspace{0.3cm}
 The paper is organized as follows. In Sec.~\ref{The action} we introduce the action of quadratic gravity we consider in this work and the corresponding  $3+1$ formalism, where space and time are separated. Afterwards, in Sec.~\ref{Canonical quadratic gravity}  Ostrogradsky's canonical method is non-perturbatively applied to quadratic gravity   to define the canonical variables and the Hamiltonian. The Dirac-Pauli quantization is applied to quadratic gravity in Sec.~\ref{Euclidean path integral and Dirac-Pauli variables} to find a consistent Euclidean path integral. At the beginning of this section a review of the DP quantization and consequent norm operators is provided to render the paper self-contained. The unitarity of the theory is discussed there in the most general terms.  In Sec.~\ref{Euclidean path integral and Dirac-Pauli variables} the Euclidean path integrals for the transition amplitudes and Green's functions are studied first in a scalar-field four-derivative interacting model and then in quadratic gravity. In the same section the general covariance of the theory is also discussed. Sec.~\ref{The quadratic gravity effective action} is devoted to the construction of the (quantum) effective action in quadratic gravity, which is then used to study the classical limit of the theory in Sec.~\ref{The classical limit}.  Sec.~\ref{Quantum cosmology}  is an application of the previous results to quantum cosmology. Finally, in Sec.~\ref{conclusions} a detailed summary of the paper and conclusions are offered. The paper also includes two appendices, which give further details.

Regarding notation: since we will consider the classical limit we will leave $\hbar$ explicit in this paper. However, we will use units where the speed of light equals 1.

\section{The action and the $3+1$ formalism}\label{The action}

  The action of quadratic gravity we consider is 
\be S=\int d^4x   \, \Lag, \qquad \mbox{with}\quad  \mathscr{L} = -\sqrt{-g}\left(\frac{\alpha}{2} W^2+\beta R^2 + \gamma R + \lambda\right)\label{TotAction}. \ee
Here $\alpha, \beta, \gamma$ and $\lambda$ are real coefficients, $g$ is the determinant of the metric, $R$ is the Ricci scalar and $W^2\equiv W_{\mu\nu\rho\sigma}W^{\mu\nu\rho\sigma}$ is the ``square" of the Weyl tensor $W_{\mu\nu\rho\sigma}$. So, in terms of the Riemann tensor $R_{\mu\nu\rho\sigma}$ and Ricci tensor $R_{\mu\nu}$
 \beq  \frac12 W^2  =  \frac12 R_{\mu\nu\rho\sigma}R^{\mu\nu\rho\sigma} - R_{\mu\nu}R^{\mu\nu} + \frac16 R^2, \eeq
 or equivalently
 \be \frac{W^2}{2}= R_{\mu\nu}R^{\mu\nu}  - \frac{R^2}{3} + \frac{G}{2}.  \ee 
 The quantity $G$ is the topological Gauss-Bonnet term.
For simplicity we do not add to $S$ possible boundary terms that are also quadratic in the curvature.
 
 One of our goals is to provide a Hamiltonian formulation of the theory. We thus adopt a $3+1$ formalism~\cite{DeWitt:1967yk,Arnowitt:1959ah,Arnowitt:1962hi}. Such a formalism has been applied to quadratic gravity in~\cite{Buchbinder-Lyahovich:1987,Buchbinder:1991ne,Buchbinder:1992rb,Kluson:2013hza}. Unlike in these works, we shall start here with a gauge condition on the metric components to avoid dealing with complicated constraints. In this respect, our approach resembles the one used  in the context of ordinary Yang-Mills theory (see e.g.~\cite{Weinberg:1996kr}). Our gauge choice is the simple Gauss coordinate system,  namely normal coordinates (zero shift) and geodesic slicing (unit lapse function):
 \be ds^2 = g_{ij}(x)dx^idx^j-dt^2, \label{GaussCo} \ee
 where latin indices are the spacial ones and $t\equiv x^0$ is the time coordinate. General covariance will be made manifest in Sec.~\ref{Transition to generic coordinate systems}.
 The non-vanishing components of the Levi-Civita connection 
 \be \Gamma_{\mu\nu}^{\rho}= \frac12 g^{\rho\sigma}(\partial_\mu g_{\sigma\nu}+\partial_\nu g_{\sigma\mu}-\partial_\sigma g_{\mu\nu})\ee
in Gauss coordinates  are then
 \be \Gamma^0_{ij}=\Gamma^0_{ji}= \frac12 \dot g_{ij}\equiv -K_{ij}, \quad \Gamma^l_{ij}, \quad \Gamma^l_{0j}=\Gamma^l_{j0}= \frac12 g^{lm}\dot g_{mj} \equiv -K^l_{~j}, \label{Gamma31} \ee
where a dot represents a derivative with respect to $t$.

The action $S$ can be expressed in terms of $g_{ij}$, $K_{ij}$ and $\dot K_{ij}$  through the following formul\ae, which can be derived with the help of those reported in Appendix~\ref{Curvature tensors}:
\bea R &=& R^\3- 2 g^{ij} \dot K_{ij}-3K_{ij}K^{ij}+K^2, \label{R31}\\
R_{\mu\nu}R^{\mu\nu} &=& \left[g^{ij} (\dot K_{ij}+K_{il}K^l_{~j})\right]^2-2(D_iK-D^jK_{ji})(D^iK-D^lK_l^{~i}) \nonumber\\ &&+(~R^\3_{ij}-\dot K_{ij}-2K_{il}K^l_{~j}+KK_{ij})g^{il}g^{jm}(~R^\3_{lm}-\dot K_{lm}-2K_{lp}K^p_{~m}+KK_{lm}), \label{Rmunu31}\\
R_{\mu\nu\rho\sigma}R^{\mu\nu\rho\sigma} &=& 4(\dot K_{ij}+K_{il}K^l_{~j})g^{il}g^{jm}(\dot K_{lm}+K_{lp}K^p_{~m})-4(D^jK_{il}-D_lK^j_{~i})(D_jK^{il}-D^lK_j^{~i}) \nonumber\\
&&+(~R^\3_{ijlm}+K_{il}K_{jm}-K_{jl}K_{im})(~R^{\hspace{-0.4cm}3 \hspace{0.3cm}ijlm}+K^{il}K^{jm}-K^{jl}K^{im}).\label{Rmunurhosigma31}
 \eea
Here \,$R^\3_{ijlm}$, \,$R^\3_{ij}$, \,$R^\3$ and $D_i$ are, respectively, the three-dimensional Riemann tensor, Ricci tensor, Ricci scalar and covariant derivative built with the three-dimensional metric $g_{ij}$, the inverse three-dimensional metric is represented as usual by $g^{ij}$ (we lower and raise the latin indices with $g_{ij}$ and $g^{ij}$, respectively) and $K\equiv K_i^{~i}$. 

The time-reversal invariance of $\mathscr{L}$ is manifest in Eqs.~(\ref{R31}),~(\ref{Rmunu31}),~(\ref{Rmunurhosigma31}): the time-reversal invariant quantity $\dot K_{ij}$ also appears linearly, while $K_{ij}\equiv -\dot g_{ij}/2$, which changes sign under time reversal, only appears quadratically. The time-reversal invariant metric $g_{ij}$ even enters non polynomially.

\section{Canonical quadratic gravity}\label{Canonical quadratic gravity}

The action $S$ is a functional of $g_{ij}$, $\dot g_{ij}$ and $\ddot g_{ij}$ or, equivalently, of $g_{ij}$, $K_{ij}$ and $\dot K_{ij}$, where $K_{ij}$ has been defined in Eq.~(\ref{Gamma31}). We can thus use Ostrogradsky's canonical method~\cite{ostro} (see e.g.~Ref.~\cite{Salvio:2018crh} for an introduction and references to previous works). 

In this formalism one can treat $g_{ij}$ and $K_{ij}$ as independent canonical coordinates and then defines the momenta $\pi^{ij}$ and $P^{ij}$ conjugate  to  $g_{ij}$ and $K_{ij}$, respectively, as follows:\footnote{The derivatives in~(\ref{MomDef}) are taken as if the variables with $i$ and $j$ exchanged were independent. For example, 
\be \frac{\partial }{\partial \dot K_{ij}}(g^{lm}\dot K_{lm}) = g^{ij}.\ee
 The symmetry in the exchange of $i$ and $j$ is then reflected in the symmetry $\pi^{ij}=\pi^{ji}$ and $P^{ij}=P^{ji}$.}
\be \pi^{ij} \equiv \frac{\partial \Lag}{\partial \dot g_{ij}} -\frac{d}{dt}\frac{\partial \Lag}{\partial \ddot g_{ij}}, \quad P^{ij} \equiv \frac{\partial \Lag}{\partial \dot K_{ij}} -\frac{d}{dt}\frac{\partial \Lag}{\partial \ddot K_{ij}} \label{MomDef}\ee
Since $\Lag$ is independent of $\ddot K_{ij}$, 
\be P^{ij} = \frac{\partial \Lag}{\partial \dot K_{ij}} \label{Pij}. \ee
By using Eqs.~(\ref{TotAction}),~(\ref{R31}),~(\ref{Rmunu31}) and~(\ref{Rmunurhosigma31}) one obtains
\bea P^{ij} = -2 \sqrt{-g}\left\{G^{ijlm}\dot K_{lm} + \alpha K^{ij} K -\alpha g^{ij} K_{lm} K^{lm}+\alpha~\,R^{\hspace{-0.4cm}3\hspace{0.3cm}ij}\right. \nonumber \\
\left.-g^{ij}\left[\left(\frac{\alpha}3+2\beta\right)(~R^\3-3K_{lm}K^{lm}+K^2)+\gamma\right]\right\}, \label{Pijsol} \eea
where 
\be G^{ijlm} \equiv \alpha\,\frac{g^{il}g^{jm}+g^{im}g^{jl}}2 +\left(4\beta-\frac{\alpha}3\right) g^{ij}g^{lm}. \ee
It is easy to check that, whenever $\alpha\neq0$ and $\beta\neq 0$, 
\be \Gamma_{pqij} \equiv   \frac1{\alpha} \frac{g_{pi}g_{qj}+g_{pj}g_{qi}}2 - \frac{4\beta-\alpha/3}{12\beta\alpha} g_{pq}g_{ij} \ee
is the inverse of $G^{ijlm}$ in the sense that 
\be \Gamma_{pqij}G^{ijlm} = \frac12 (\delta_p^l\delta_q^m+\delta_p^m\delta_q^l),\ee
so we can express $\dot K_{ij}$ in terms of $P^{ij}$, $g_{ij}$ and $K_{ij}$. Inserting this expression in $\Lag$ one finds a functional of $g_{ij}$, $K_{ij}$ and $P^{ij}$ with no dependence on $\pi^{ij}$. 

Also in Ostrogradsky's canonical method, one makes  use of the standard definition of the classical Hamiltonian density: 
\be {\cal H} \equiv \pi^{ij} \dot g_{ij} + P^{ij} \dot K_{ij} - \Lag. \label{HdenQG}\ee
As usual, the spatial integral of ${\cal H}$ (namely, the Hamiltonian) is conserved, i.e.~time independent. 
Unlike $\Lag$, the Hamiltonian density  does depend on $\pi^{ij}$, as well as $g_{ij}$, $K_{ij}$ and $P^{ij}$, but the only dependence on $\pi^{ij}$ is given by the term
\be \pi^{ij} \dot g_{ij} = - 2 \pi^{ij} K_{ij}. \label{UnbTerm}\ee
As mentioned in the introduction, the {\it classical} Hamiltonian is, therefore, unbounded from below, but the runaway rates in classical quadratic gravity can be slow enough to ensure compatibility with observations in our observable patch, while other highly inhomogeneous and anisotropic patches are not compatible with observers thanks to Ostrogradsky's instabilities (leading to an explanation of the nearly homogeneity and isotropy of our universe).  In Sec.~\ref{Quantum cosmology} we will discuss the quantum counterpart of these classical findings.

\section{Euclidean path integral and Dirac-Pauli variables}\label{Euclidean path integral and Dirac-Pauli variables}

As usual the dynamics of the quantum theory will be encoded in a time-evolution operator $U(t)\equiv\exp(-i H t/\hbar)$, where $H$ is the Hamiltonian operator. 
As usual in a field theory, we seek here an equivalent formulation in terms of a path integral. 

The Euclidean path integral, followed by the analytic continuation to real time, provides us with a rigorous non-perturbative definition of our physical theories. It is thus important to construct a consistent Euclidean path integral for quadratic gravity too. In the Euclidean theory the real time $t$ is substituted by $-i\tau$, where $\tau$, called the imaginary time, is treated as a real variable. The real-time (Lorentzian) theory is then {\it defined} by the analytic continuation $\tau\to it$ of the resulting theory.

A first problem one faces to achieve this goal is the presence of the term~(\ref{UnbTerm}) in ${\cal H}$: if one performs an ordinary quantization where the canonical coordinates $g_{ij}$ and $K_{ij}$ (and their conjugate momenta $\pi^{ij}$ and $P^{ij}$) have a real spectrum the Euclidean path integral over $\pi^{ij}$ diverges: this is because in the Euclidean path integral the Hamiltonian appears without an overall $i$ in the exponent of the path integrand. 

A solution to this problem is the existence of an alternative quantization for the canonical variables, where these operators have instead purely immaginary eigenvalues. We will refer to this quantization as the Dirac-Pauli (DP) one~\cite{Salvio:2015gsi,Salvio:2018crh,Salvio:2020axm}, as it was first discussed by Pauli~\cite{Pauli} elaborating on a previous work by Dirac~\cite{Dirac}. The existence of this alternative quantization is due to the fact that an observable $A$ in quantum mechanics can be more generally defined as a physical quantity for which there is a {\it complete} set of states with a well defined value of $A$\cite{Salvio:2019wcp,Salvio:2020axm}. As a result of this more general definition, $A$ can always be described by a normal operator with respect to an appropriately defined positive-definite inner product~\cite{Salvio:2019wcp,Salvio:2020axm} with eigenvectors given by the mentioned complete set of states. This extends the usual definition of observables as $A$ can have complex eigenvalues. This is not so strange as we can construct normal non-Hermitian observables even in  text-book quantum mechanics: for example, the position of a particle in a plane $\{x, y\}$ can always be described by a normal operator $\hat x+i\hat y$ with complex eigenvalues. With this definition of observable (which is adopted here) there are two possibilities to quantize a canonical variable, either the ordinary one or the DP one. Now, if we adopt the ordinary quantization for $\pi^{ij}$ and the DP one for $K_{ij}$ the term in~(\ref{UnbTerm}) will produce a harmless purely imaginary term in the Euclidean path integral and the previously mentioned problem is solved.

To proceed it is now necessary to review some of the properties of this generalized quantization procedure. First a DP canonical variable $\hat x$ satisfies by definition $\hat x |x\rangle = ix|x\rangle$ for a complete set of states $|x\rangle$ and real $x$. The operator $\hat x$ can be considered as a Hermitian operator with respect to an indefinite inner product for which $\langle x'|x\rangle = \delta(x+x')$ rather than $\delta(x-x')$~\cite{Salvio:2018crh}. For generic states $|\psi\rangle$ and $|\chi\rangle$ such inner product is denoted $\langle \chi|\psi\rangle$ and the Hermitian conjugate with respect to this inner product of a generic linear operator $O$ is denoted $O^\dagger$. Calling $\eta$ the linear operator such that $\eta|x\rangle = |-x\rangle$, it is then clear that $\langle \chi|\eta|\psi\rangle$ defines a positive-definite inner product. The completeness relation then reads
\be \int dx|x\rangle\langle x|\eta =1,  \qquad \int|x\rangle\langle x| =\eta, \ee
where in the second equality we used $\eta^2 =1$. Also it is easy to show $\eta^\dagger = \eta$. If we call $O^\dagger_\eta$ the Hermitian conjugate of a generic linear operator $O$ with respect to the positive-definite inner product, $\langle \chi|\eta|\psi\rangle$, one finds $O^\dagger_\eta = \eta O^\dagger \eta$.

More generally, for any observable $A$ it is always possible to define a linear operator $P_A$ (satisfying $P_A^\dagger = P_A$) such that for the  complete set of eigenstates $|a\rangle$ of $A$ with eigenvalue $\alpha_a$ one has $\langle a'|P_A|a\rangle=\delta_{a'a}$ and thus  $\langle \chi|\psi\rangle_A\equiv\langle \chi|P_A|\psi\rangle$ is a positive-definite inner product~\cite{Salvio:2019wcp,Salvio:2020axm} (when $A$ is a DP canonical variable one has $P_A = \eta$). For this reason we refer to $P_A$  as a norm operator. Adopting the standard frequency definition of probability as in real physics experiments, one finds that it is this inner product that must be used in the Born rule to define the probability $p_a$ that a measurement of $A$ has the outcome $\alpha_a$~\cite{Strumia:2017dvt,Salvio:2019wcp,Salvio:2020axm}:  
 \be p_a = \frac{|\langle a|\psi\rangle_A|^2}{\langle\psi|\psi\rangle_A}. \label{ProbNorm}\ee
 Since this satisfies for any $A$ the conditions~\cite{Salvio:2019wcp,Salvio:2020axm}
 \be p_a\geq 0, \qquad \sum_a p_a = 1, \label{GenUnit} \ee
 where the label $a$ ranges over all possible values, the theory is unitary. 
 
 Another property that is necessary to recall is that the conjugate momentum $\hat p$ of a DP canonical coordinate $\hat x$ (satisfying the canonical commutator $[\hat x, \hat p] = i\hbar$) is itself a DP variable, namely $\hat p|p\rangle = i p |p \rangle$ for a complete set of states $|p\rangle$ and $p$ real and~\cite{Salvio:2015gsi,Salvio:2018crh} 
 \be \langle x|p\rangle = \frac{e^{i p x/\hbar}}{\sqrt{2\pi\hbar}}.\ee
 Therefore, the quantization that was mentioned before (the ordinary one for $\pi^{ij}$ and the DP one for $K_{ij}$) tells us that the pairs of conjugated variables $\{g_{ij}, \pi^{ij}\}$ and $\{K_{ij}, P^{ij}\}$ are respectively ordinary variables and DP ones. This agrees with the principle proposed in~\cite{Salvio:2015gsi} in the simpler Pais-Uhlenbeck model~\cite{Pais}: coordinates that are invariant under time reversal  (like $g_{ij}$) should be quantized in the standard way, while coordinates that are odd under time reversal (like $K_{ij} \equiv -\dot g_{ij}/2$) should be DP quantized\footnote{After quantization we write $Tg_{ij}T^{-1} = g_{ij}$ and $T\dot g_{ij}T^{-1} = -\dot g_{ij}$, where $T$ is the antilinear time-reversal operator.}. One might worry that this quantization will have the disastrous  implication that the theory predicts an imaginary value for the time derivative of the metric;  this does not happen, as will be confirmed in Secs.~\ref{The quadratic gravity effective action} and~\ref{The classical limit}, because the expectation values of all types of canonical coordinates (both ordinary and DP ones) are real~\cite{Salvio:2020axm}.

 It remains to be shown  that this quantization leads to a well-defined Euclidean path integral as anticipated. This will be done in the following subsection.

 \subsection{Euclidean path integral for transition amplitudes and T-products} \label{EPTTp}
 
 In this section it is convenient to treat  a more general framework than quadratic gravity. Let us consider an arbitrary number of ordinary canonical variables $q_1, ..., q_n$ and Dirac-Pauli variables $\bar q_1, ..., \bar q_m$. A state with definite canonical coordinates is denoted here~\cite{Salvio:2018crh}
 \be 
 |q\rangle = |q_1, ..., q_n, \bar q_1, ..., \bar q_m \rangle\ee
 and the corresponding state with definite conjugate momenta is 
  \be 
 |p\rangle = |p_1, ..., p_n, \bar p_1, ..., \bar p_m \rangle.\ee
 
 We would like to evaluate the matrix element of the Euclidean time evolution operator, $\exp(-(\tau_f-\tau_i) H/\hbar)$ with the imaginary initial and final times, $\tau_i$ and $\tau_f$ respectively, between arbitrary states of definite canonical coordinates $|q_i\rangle$ and $|q_f\rangle$:
 \be \langle q_f|\eta\exp(-(\tau_f-\tau_i) H/\hbar) |q_i\rangle. \ee
 Indeed, the knowledge of this object for arbitrary $q_i$ and $q_f$ is equivalent to knowing the imaginary-time dynamics of the system. The real-time dynamics can then be defined as an analytic continuation from imaginary to real time. In the expression above we have inserted the $\eta$ operator so that the matrix element is taken with the positive-definite inner product $\langle \chi|\psi\rangle_q\equiv\langle \chi|\eta|\psi\rangle$. The matrix element without the $\eta$ insertion provides an object  with the same amount of information because $\eta$ by definition switches the sign of all DP variables and leaves the ordinary variables unchanged, i.e.
 \be \eta |q\rangle \equiv |q_\eta\rangle \equiv |q_1, ..., q_n, -\bar q_1, ..., -\bar q_m \rangle .\ee
 So the desired matrix element can be written as follows:
 \be \langle q_f|\eta e^{-(\tau_f-\tau_i) H/\hbar} |q_i\rangle = \langle q_{f\eta},\tau_f|q_i,\tau_i\rangle,  \label{EuMatEl}\ee
 having defined $|q,\tau\rangle\equiv \exp(\tau H/\hbar)|q\rangle$ and $\langle q,\tau|\equiv \langle q|\exp(-\tau H/\hbar)$. After an analytic continuation to real time, these matrix elements allows us to compute, among other things, the probability density $\rho(q_f)$ that the system is found in the state $|q_f\rangle$ after a time $t_f-t_i$ starting from a generic state $|\psi_0\rangle = \int dq\,\psi_0(q) |q\rangle$ by using Eq.~(\ref{ProbNorm}) with $A =q$:
 \be \hspace{-.65cm}\rho(q_f) = \frac{|\langle q_{f\eta}|U(t_f-t_i)|\psi_0\rangle|^2}{\langle \psi_0|U^\dagger(t_f-t_i) \eta U(t_f-t_i)|\psi_0\rangle} =  \frac{|\langle q_{f\eta}|U(t_f-t_i)|\psi_0\rangle|^2}{\int dq_\eta |\langle q_\eta|U(t_f-t_i)|\psi_0\rangle|^2}, \label{probDens} \ee
 where  in the second step we have changed the name of the integration variable, $q\to q_\eta$. From Eq.~(\ref{probDens}) it is clear that the probability density $\rho(q_f)$ satisfies the {\it unitarity conditions}
 \be \boxed{\rho(q_f)\geq 0,~~\mbox{(for any \,$q_f$)}, \quad \int dq_f\, \rho(q_f) =1} \label{unitaryrho} \ee  
 as a consequence of the general unitarity expressed by~(\ref{GenUnit}).

We actually consider an even more general object than~(\ref{EuMatEl}), where a generic time-ordered product of ordinary canonical coordinates of the form T$\hat q_{i_1}(\tau_1)\hat q_{i_2}(\tau_2) ...$ is inserted:
\be \langle q_{f\eta},\tau_f|\mbox{T}\hat q_{i_1}(\tau_1)\hat q_{i_2}(\tau_2)... |q_i,\tau_i\rangle, \ee 
where $\hat q(\tau) \equiv \exp(H\tau/\hbar)\hat q\exp(-H\tau/\hbar)$. This generalization will allow us in Sec.~\ref{EPTTpp} to construct the Euclidean path integral for Green's functions of the metric $g_{ij}$, which, as we have seen, will be quantized as an ordinary canonical coordinate.  
By following steps similar to those  in the absence of such insertion performed in Refs.~\cite{Salvio:2015gsi,Salvio:2018crh}, we obtain the path-integral representation:
\be   \langle q_{f\eta},\tau_f|\mbox{T}\hat q_{i_1}(\tau_1)\hat q_{i_2}(\tau_2)... |q_i,\tau_i\rangle  = \int^{q(\tau_f) = q_f}_{q(\tau_i) = q_i}  \delta q \delta p~q_{i_1}(\tau_1)q_{i_2}(\tau_2)...\exp\left( \frac1{\hbar}\int_{\tau_i}^{\tau_f} d\tau (i pq'-\bar H(q,p))\right),  \label{PIEuclidean}\ee 
where 
\be \bar H(q,p) \equiv \frac{\langle p| H | q\rangle}{ \langle p | q\rangle} \label{Hbardef}\ee 
and the functional integration measure is given by 
\be \delta q = \prod_{\tau} dq(\tau), \quad \delta p =  \prod_{\tau'}  \frac{dp(\tau)}{2\pi\hbar}. \label{timeEprod}\ee
The products above are extended to all values of $\tau\in[\tau_i,\tau_f)$ for momenta $p$ and for all values of $\tau\in(\tau_i,\tau_f)$ for coordinates $q$, as the values of $q$ at both time boundaries are fixed (we put a prime in the second product to recall this different range).
Like in any quantum theory, this measure should be thought of as the zero lattice-spacing limit of an integration measure over a finite number of variables obtained by discretizing time (and all spatial dimensions in the case of field theory).
As usual we can write
\be \langle q_{f\eta},\tau_f|\mbox{T}\hat q_{i_1}(\tau_1)\hat q_{i_2}(\tau_2)... |q_i,\tau_i\rangle  = \left.\frac{\delta \langle q_{f\eta},\tau_f|q_i,\tau_i\rangle^J}{\delta J_{i_1}(\tau_1)\delta J_{i_2}(\tau_2) ... }\right|_{J=0},  \ee
where $\langle q_{f\eta},\tau_f|q_i,\tau_i\rangle^J$ is the generating functional 
\be \boxed{\langle q_{f\eta},\tau_f|q_i,\tau_i\rangle^J \equiv \int^{q(\tau_f) = q_f}_{q(\tau_i) = q_i} \delta q \delta p~\exp\left( \frac1{\hbar}\int_{\tau_i}^{\tau_f} d\tau (i pq'-\bar H(q,p)+\hbar J(\tau)q(\tau))\right)}\label{PIEuclideanG}\ee

It is important to note that here the expression $q(\tau_i) = q_i, \, q(\tau_f) = q_f$ appearing on the functional integral symbol represents boundary conditions for both the ordinary and DP coordinates, both at initial time $\tau_i$
\be q_1(\tau_i) = q_{1i}, ..., q_n(\tau_i) = q_{ni}, \bar q_1(\tau_i) =\bar q_{1i}, ..., \bar q_m(\tau_i) = \bar q_{mi}, \ee 
and final time $\tau_f$
\be q_1(\tau_f) = q_{1f}, ..., q_n(\tau_f) = q_{nf}, \bar q_1(\tau_f) =\bar q_{1f}, ..., \bar q_m(\tau_f) =\bar  q_{mf}. \ee 
Another important aspect is that in the presence of DP variables $\bar H(q,p)$ does not generically coincide with the classical Hamiltonian (the space integral of the classical ${\cal H}$ given in~(\ref{HdenQG}) in the case of quadratic gravity) because the eigenvalues of the DP variables are imaginary. So, in the limit $\hbar\to 0$, where all operators commute, $\bar H(q,p)$ is the analytic continuation of the classical Hamiltonian to imaginary values of the DP variables: $\bar q_1 \to i \bar q_1$, ..., $\bar q_m \to i \bar q_m$, $\bar p_1 \to -i \bar p_1$, $\bar p_m \to -i \bar p_m$. The DP coordinates are rotated anticlockwise in the complex plane while the DP conjugate momenta are rotated clockwise because in the definition of $\bar H$ we are adopting, Eq.~(\ref{Hbardef}), the coordinate eigenstates appear on the right, while the momentum eigenstates are on the left. 

The path integral in~(\ref{PIEuclidean}), or equivalently that in~(\ref{PIEuclideanG}), is not guaranteed to converge for any theory and for any choice of quantization. As we have anticipated at the beginning of Sec.~\ref{Euclidean path integral and Dirac-Pauli variables}, if we use the ordinary quantization for both  $\{g_{ij}, \pi^{ij}\}$ and $\{K_{ij}, P^{ij}\}$ in quadratic gravity, $\bar H$ will have a term $- 2 \pi^{ij} K_{ij}$, where now $\pi^{ij}$ and $K_{ij}$ are real integration variables in the path integral. Since this is the only term where $\pi^{ij}$ appears, one would obtain an exponential divergence of the path integral for this quantization.  In general one can only accept  quantizations (and parameter values) leading to an $\bar H(q,p)$ that guarantees the convergence of the 
Euclidean path integral. 

Now, to specify the quantization completely we  still have to uniquely define the operator $H=\int d^3x\,{\cal H}$, without ambiguities. The analysis of Sec.~\ref{Canonical quadratic gravity} was purely classical and, as well known, classical physics cannot uniquely determine quantum physics. We now define the operator $H$ as the function of the canonical variable operators obtained by  first putting all canonical momenta $c$-numbers on the left of all  canonical coordinate $c$-numbers in the classical Hamiltonian and then substituting these $c$-numbers with the corresponding operators.  In this way, $\bar H$ in~(\ref{Hbardef}) can be easily computed and coincides with the classical Hamiltonian in the absence of DP variables. In the presence of DP variables, $\bar H(q,p)$ is, now for any $\hbar$, the analytic continuation of the classical Hamiltonian to imaginary values of the DP variables: $\bar q_1 \to i \bar q_1$, ..., $\bar q_m \to i \bar q_m$, $\bar p_1 \to -i \bar p_1$, $\bar p_m \to -i \bar p_m$.  

Note that the quantum Hamiltonian $H$ of quadratic gravity is time-reversal invariant, $T H T^{-1} = H$. The quantum theory has inherited this symmetry property from the time-reversal invariance of the classical theory discussed in Sec.~\ref{The action}.

\subsubsection*{Application to a four-derivative scalar-field theory}

To understand how this formalism works in a simple context, we first consider a scalar-field theory with Lagrangian density
\be \Lag = -\frac12 \partial_\mu\varphi \partial^\mu\varphi - \frac{c_4}{2} (\Box \varphi)^2 - V(\varphi), \label{ScalLag}\ee 
where $\Box \equiv \partial_\mu\partial^\mu$, $c_4$ is a real coefficient governing the size of the four-derivative term and $V$ is some potential density. In the $3+1$ formalism we have 
 \be \Lag = \frac12 \dot\varphi^2 -\frac12(\nabla\varphi)^2- \frac{c_4}{2} (\ddot\varphi-\nabla^2\varphi)^2 - V(\varphi), \label{ScalLag2} \ee 
 where $(\nabla\varphi)^2\equiv \partial_i\varphi\partial_i\varphi$ and $\nabla^2\varphi\equiv \partial_i\partial_i\varphi$.
 
 Like we did for quadratic gravity, we introduce another canonical coordinate given by the time derivative of the dynamical field $\varphi_2\equiv\dot\varphi$. Following Ostrogradsky's canonical method the momenta conjugate to $\varphi_1\equiv \varphi$ and $\varphi_2\equiv\dot\varphi$ are, respectively, 
 \be \pi_1 = \frac{\partial\Lag}{\partial \dot\varphi_1} -\frac{d}{dt} \frac{\partial\Lag}{\partial \ddot\varphi_1} = \dot\varphi + c_4(\dddot \varphi-\nabla^2\dot\varphi), \quad \pi_2 = \frac{\partial\Lag}{\partial \dot\varphi_2} -\frac{d}{dt} \frac{\partial\Lag}{\partial \ddot\varphi_2} = c_4(\nabla^2\varphi-\ddot\varphi). \label{pi12def}\ee
 The last equation can be used to express $\ddot\varphi$ in terms of the canonical variables,  $\ddot\varphi = \nabla^2\varphi_1-\pi_2/c_4$. 
 
 The Hamiltonian density with the ordering specified above (all canonical momenta to the left of all canonical coordinates\footnote{One can note, however, that this specification is irrelevant for the scalar theory discussed here, unlike quadratic gravity, because the momenta $\pi_1$ and $\pi_2$ only multiply the other coordinates or functionals of the other coordinates, $\varphi_2$ and $\varphi_1$, respectively, and $[\pi_1, \varphi_2]=[\pi_2,\varphi_1]=0$.}) then reads
 \bea {\cal H} = \pi_1 \dot \varphi_1 +\pi_2 \dot\varphi_2 - \Lag  = \pi_1\varphi_2-\frac{\pi_2^2}{2c_4}+\pi_2\nabla^2\varphi_1-\frac12 \varphi_2^2+\frac12(\nabla\varphi_1)^2 + V(\varphi_1).\label{ScalHam}\eea
 This simple model presents some key features of quadratic gravity. The first conjugate momentum $\pi_1$ only appears in the term $\pi_1\varphi_2$ and using the ordinary quantization for both pairs of canonical variables would lead to a divergent $\pi_1$ integration in the Euclidean path integral. Moreover, the classical theory is time-reversal invariant, as clear from Eqs.~(\ref
 {ScalLag2}) and~(\ref{pi12def})-(\ref{ScalHam}). Like for quadratic gravity (and the even simpler Pais-Uhlenbeck model, see~\cite{Salvio:2015gsi,Salvio:2018crh}), we then use the ordinary quantization for the $T$-even coordinate $\varphi_1\equiv \varphi$ (i.e.~$T \varphi T^{-1}=\varphi$) and the DP one for the $T$-odd coordinate $\varphi_2\equiv\dot\varphi$ (i.e.~$T \dot \varphi T^{-1}=-\dot\varphi$). This implies that $\pi_1$ and $\pi_2$ should be quantized in the ordinary and DP way, respectively.  The operator $H$ is then just $\int d^3x\,{\cal H}$ with ${\cal H}$ given here in~(\ref{ScalHam}) and with the quantization just specified. Note that, like for quadratic gravity, $H$ is time-reversal invariant, $THT^{-1}=H$, as a consequence of the time-reversal invariance of the classical theory. Moreover, the quantity $\bar H$ defined in~(\ref{Hbardef}) is here 
 \be \hspace{-0.2cm}\bar H(\varphi_1,\varphi_2,\pi_1,\pi_2) = \int d^3x\, {\cal\bar H}, \quad {\cal\bar H}= \left(i\pi_1\varphi_2+\frac{\pi_2^2}{2c_4}-i\pi_2\nabla^2\varphi_1+\frac12 \varphi_2^2+\frac12(\nabla\varphi_1)^2 + V(\varphi_1)\right),  \label{barHSF}\ee 
 where now $\{\varphi_1,\pi_1\}$ and $\{\varphi_2,\pi_2\}$ are $c$-numbers because are integration variables in the path integral.
 In Table~\ref{DPvsOrd} the type of quantization for the canonical variables for this scalar theory and quadratic gravity is summarized.
  \begin{table}[t]
\centering
\begin{tabular}{ |p{4cm}|p{4cm}|p{4cm}| }
\hline
\rowcolor{red!80!green!40!yellow!10}theory& ordinary variables & DP variables \\
\hline
 four-derivative scalar & $\{\varphi_1\equiv\varphi,\pi_1\}$ & $\{\varphi_2\equiv\dot\varphi,\pi_2\}$ \\
quadratic gravity &  $\{g_{ij},\pi^{ij}\}$   & $\{K_{ij}\equiv -\dot g_{ij}/2,P^{ij}\}$ \\
\hline
\end{tabular}
\caption{\it Quantization for the canonical variables in the scalar theory with Lagrangian density~(\ref{ScalLag}) and in quadratic gravity. The conjugate momenta are defined in~(\ref{pi12def}) and~(\ref{MomDef}).}
\label{DPvsOrd}
\end{table}

 The previously dangerous $\pi_1\varphi_2$ has now become a harmless imaginary term $i\pi_1\varphi_2$. The integration over $\pi_1$ and $\varphi_2$ in~(\ref{PIEuclideanG}) then sets $\varphi_2 = \varphi'$ and multiplies the integration measure by $\prod_{x_E'} \Delta \tau^{-1}$, where $\Delta \tau$ is the Euclidean time element (which becomes $d\tau$ in the zero lattice-spacing limit) and the product is over all Euclidean spacetime points $x_E$ with imaginary time $\tau \in[\tau_i,\tau_f)$. We put a prime on $x_E$ to distinguish this product with that for canonical coordinates: as we have discussed below~(\ref{timeEprod}), for canonical coordinates, like $\varphi$, the product over the imaginary time ranges over $\tau \in(\tau_i,\tau_f)$ while for conjugate momenta over $\tau \in[\tau_i,\tau_f)$. The integration over $\pi_2$ then reads
 \be \int  \exp\left(\int_{\tau_i}^{\tau_f} d\tau\int d^3x\left[ i\pi_2(\nabla^2\varphi+\varphi'') - \frac{\pi_2^2}{2c_4}\right]/\hbar \right) \prod_{x_E'} \frac{\Delta V_3d\pi_2(x_E)}{2\pi\hbar}, \label{intpi2} \ee
 where $\Delta V_3$ is the spatial volume element (which becomes $d^3x$ in the zero lattice-spacing limit). This factor of $\Delta V_3$ appears in the measure because we have defined $\pi_1$ and $\pi_2$ through the derivatives of the Lagrangian density $\Lag$ rather than the Lagrangian. 
The integral in~(\ref{intpi2}) converges only for $c_4>0$, which we assume, and, using a standard stationary point method, gives
\be C' \exp\left(-\frac{c_4}{2} \int_{\tau_i}^{\tau_f} d\tau\int d^3x(\nabla^2\varphi+\varphi'')^2\right), \quad\mbox{with} \quad C' = \prod_{x_E'}\sqrt{\frac{c_4\Delta V_3}{2\pi\hbar\Delta\tau}}. \ee 
So we arrive at the Lagrangian version of the Euclidean path integral 
\be \langle q_{f\eta},\tau_f|q_i,\tau_i\rangle^J  =\int^{q(\tau_f) = q_f}_{q(\tau_i) = q_i}  \, C\delta \varphi \, \exp\left(-S_E/\hbar  + \int_{\tau_i}^{\tau_f} d\tau\int d^3x J\varphi\right), \label{PIEucvarphi}\ee
where\footnote{For a standard four-dimensionally cubic lattice $\Delta \tau =a$ and $\Delta V_3=a^3$, where $a$ here is the lattice spacing, so $C = \prod_{x'_E} \sqrt{\frac{c_4}{2\pi\hbar}}$.}
\be C = \prod_{x_E'}\sqrt{\frac{c_4\Delta V_3}{2\pi\hbar\Delta\tau^3}},\ee
$S_E$ is the Euclidean action 
\be S_E = \int_{\tau_i}^{\tau_f} d\tau\int d^3x\left( \frac12 \partial_\mu\varphi\partial_\mu\varphi +\frac{c_4}{2} (\partial_\mu\partial_\mu\varphi)^2+V(\varphi)\right), \ee
and $\mu$ runs now over the Euclidean time and the spatial dimensions. In~(\ref{PIEucvarphi}) the boundary conditions can be more explicitly written as 
\be q(\tau_i) = q_i: \quad \varphi(\tau_i)=\varphi_{1i}, \varphi'(\tau_i) = \varphi_{2i} , \qquad  q(\tau_f) = q_f: \quad \varphi(\tau_f)=\varphi_{1f}, \varphi'(\tau_f) = \varphi_{2f}. \ee
We include the  factor $C$ in the definition of the integration measure in~(\ref{PIEucvarphi}), which then reads  $C\delta\varphi = C\prod_{x_E} d\varphi(x_E)$. 

One might wonder why we obtained precisely the Euclidean action at the end. While $\varphi_2\equiv \dot\varphi$ corresponds to $i\varphi'$ in Euclidean time and the eigenvalues of $\varphi_2$, beeing quantized \`a la DP, are indeed purely imaginary, $\pi_2$, which is also quantized \`a la DP,  corresponds to a real quantity in Euclidean time, see Eq.~(\ref{pi12def}). The reason why we obtained  the Euclidean action is the following. After solving the Gaussian functional integral over $\pi_2$ 
we obtain the exponential in~(\ref{PIEuclideanG})
computed  in the stationary point of its argument with respect to variations of $\pi_2$. This stationary point satisfies the equation
\be i\varphi_2' = \frac{\partial {\cal\bar H}}{\partial\pi_2}, \ee
where now the canonical variables are $c$-numbers (integration variables in the path integral). As we have seen, $\varphi_2$ is set equal to $\varphi'$ by the integration over $\pi_1$ and $\varphi_2$. So the equation above becomes
\be -\varphi'' = \frac{\partial {\cal\bar H}}{\partial(-i\pi_2)}. \ee
Since $\varphi''$ is analytically continued to $-\ddot\varphi$ in real time,  $\pi_2$ is a DP variable and the momentum eigenstates appear on the left in the definition of $\bar H$ in~(\ref{Hbardef}), the latter equation is the Euclidean version of the equation that allows us to express $\pi_2$ in terms of $\varphi$, $\dot\varphi$ and $\ddot\varphi$ in classical physics but with $\pi_2$ substituted by $-i\pi_2$, compatibly with the DP quantization. 
As a result, the exponential in~(\ref{PIEuclideanG}) becomes the exponential in~(\ref{PIEucvarphi}) with $S_E$ being the Euclidean action.

The finiteness of~(\ref{PIEucvarphi}) requires $c_4> 0$, which has indeed been assumed in the derivation, and that the potential density $V$ be bounded from below. Also,  $V$ should diverge fast enough to $+\infty$ as $\varphi\to \infty$. 
It is important to note that these conditions of convergence are on the bare parameter $c_4$ and the possible other bare parameters appearing in $V$. As in any interacting quantum field theory, the renormalized parameters are generically   different quantities, not necessarily subject to the same conditions.

Now let us consider the change of variable $\varphi(\tau) \equiv \tilde\varphi(\tilde\tau)$  in the path integral in~(\ref{PIEucvarphi}), where $\tilde\tau\equiv\tau_f+\tau_i-\tau$ and the dependence on the spatial coordinate is untouched and thus understood. This corresponds to a time reversal. Since $\varphi'(\tau) = -\tilde\varphi'(\tilde\tau)$, where a prime denotes the derivative with respect to the displayed argument, and $S_E$ is invariant with respect to time reversal, i.e. $S_E(\varphi) = S_E(\tilde\varphi)$, one obtains
\be \langle q_{f\eta},\tau_f|q_i,\tau_i\rangle^J = \langle q_{i},\tau_f|q_{f\eta},\tau_i\rangle^{\tilde J}, \label{ChangeT}\ee
where the new current $\tilde J$ is defined by $\tilde J(\tau) \equiv J(\tau_f+\tau_i-\tau)$.
So when  $J=0$
\be \langle q_f|\eta\exp(-(\tau_f-\tau_i) H/\hbar) |q_i\rangle=
\langle q_{f\eta},\tau_f|q_i,\tau_i\rangle = \langle q_{i},\tau_f|q_{f\eta},\tau_i\rangle = \langle q_i|\exp(-(\tau_f-\tau_i) H/\hbar)\eta |q_f\rangle, \ee
which means $H^\dagger = H$, or equivalently  $H^\dagger_\eta = \eta H \eta$, because the path integral in~(\ref{PIEucvarphi}) is real. In words, the quantization we have performed corresponds to having $H$ Hermitian with respect to the indefinite inner product, $\langle\chi|\psi\rangle$, defined at the beginning of Sec.~\ref{Euclidean path integral and Dirac-Pauli variables}. Nevertheless, the theory is unitary (all probabilities are non negative and sum up to one) as described around Eq.~(\ref{ProbNorm}). As we will see, the conclusion $H^\dagger=H$ holds in quadratic gravity too and in Sec.~\ref{Quantum cosmology}
 we will discuss the physical consequences.

\subsubsection*{The quadratic gravity case}

The quadratic gravity case parallels the scalar-field theory one we have just discussed, but with some important differences that we now highlight.

As already mentioned, the quantization is specified by treating $\{g_{ij},\pi^{ij}\}$ as ordinary variables, $\{K_{ij},P^{ij}\}$ as DP variables and positioning all canonical momenta on the left of all canonical coordinates in the quantum Hamiltonian. The $c$-number $\bar H$ in~(\ref{Hbardef}) is then,  for any $\hbar$, the analytic continuation of the classical Hamiltonian to imaginary values of the DP variables: $K_{ij} \to i K_{ij}$, $P^{ij} \to -i P^{ij}$. Correspondingly, ${\cal\bar H}$ is the same analytic continuation of the classical Hamiltonian  density.

The integration over $\pi^{ij}$ and $K_{ij}$ in~(\ref{PIEuclideanG}) then sets $K_{ij} = -\frac12 g_{ij}'$ and multiplies the integration measure by $\prod_{x_E'} \Delta \tau^{-6}$. The extra power of $6$ compared to the scalar-field case  is due to the fact that $\pi^{ij}$ (just as $g_{ij}$, $K_{ij}$ and $P^{ij}$) has a total of $6$ independent components at each spacetime point.  

The integral over $P^{ij}$ is, like the integral over $\pi_2$ in the scalar-field example, a Gaussian integral that can be performed with a standard stationary point method. The stationary point satisfies now the equation 
\be iK_{ij}' = \frac{\partial {\cal\bar H}}{\partial P^{ij}},\ee
which, because $K_{ij}$ has been set equal to $-\frac12 g_{ij}'$, reads
\be \frac12  g_{ij}'' = \frac{\partial {\cal\bar H}}{\partial (-iP^{ij})}. \ee
Similarly to the scalar-field case, the latter equation is the Euclidean version of the Lorentzian equation, Eq.~(\ref{Pij}), that allows us to express $P^{ij}$ in terms of $g_{\ij}$, $\dot g_{ij}$ and $\ddot g_{ij}$ in classical physics but with $P^{ij}$ substituted by $-iP^{ij}$, compatibly with the DP quantization. Note the importance of our choice of quantization in obtaining this result.
Therefore, the integral over $P^{ij}$ sets the exponent in the path integral~(\ref{PIEuclideanG}) equal to 
\be-S_E/\hbar  + \int_{\tau_i}^{\tau_f} d\tau\int d^3x \,J^{ij}g_{ij}, \label{argexpPI}\ee
where $S_E$ is the Euclidean action,
\be \boxed{S_E = \int_{\tau_i}^{\tau_f} d\tau\int d^3x\sqrt{g}\left(\frac{\alpha}{2} W^2+\beta R^2 + \gamma R + \lambda\right)} \label{SEQG}\ee
and adds the extra factor 
\be C' = \int  \exp\left(\int_{\tau_i}^{\tau_f} d\tau\int d^3x\left(  - \frac{1}{4\sqrt{g}} \eta^{pq}\Gamma_{pqij} \eta^{ij}\right)/\hbar \right) \prod_{x_E'} \prod_{l\geq m}\frac{\Delta V_3\,d\eta^{lm}(x_E)}{2\pi\hbar} \label{intetaij} \ee
due to the remaining Gaussian integral over the fluctuation $\eta^{ij}$ of $P^{ij}$ around the stationary point (recall that $P^{ij}$ is a DP variable).
In~(\ref{argexpPI}) $J^{ij}$ is the external ``current" corresponding to $g_{ij}$ in the generating functional~(\ref{PIEuclideanG}). Moreover, in~(\ref{SEQG}) the metric determinant $g$, the square of the Weyl tensor $W^2\equiv W_{\mu\nu\rho\sigma}W_{\mu\nu\rho\sigma}$ and the Ricci scalar $R$ are now computed with the Euclidean metric, which for Gauss coordinates reads
 \be ds^2_E = g_{ij}(x)dx^idx^j+d\tau^2.  \ee
 What remains to be done is to compute the factor $C'$ in~(\ref{intetaij}). This is done in Appendix~\ref{Gaussian integral over gravitational momenta}. 
 
For quadratic gravity one then finally obtains 
\be \boxed{\langle q_{f\eta},\tau_f|q_i,\tau_i\rangle^J  = \int^{q(\tau_f) = q_f}_{q(\tau_i) = q_i}  \, C\delta g \, \exp\left(-S_E/\hbar  + \int_{\tau_i}^{\tau_f} d\tau\int d^3x \, J^{ij}g_{ij}\right).} \label{PIEucQG}\ee
Here the boundary conditions can be more explicitly written as 
\bea && q(\tau_i) = q_i: \quad g_{lm}(\tau_i) = g^{(i)}_{lm},~g'_{lm}(\tau_i) = -2K_{lm}^{(i)}, \\ && q(\tau_f) = q_f: \quad g_{lm}(\tau_f) = g^{(f)}_{lm},~g'_{lm}(\tau_f) = -2K_{lm}^{(f)}, \label{gKbound} \eea
where $g^{(i,f)}_{lm}$ and $K^{(i,f)}_{lm}$ provides initial and final conditions for the metric and its time derivative; they are functions of the three-dimensional spatial coordinates.  By multiplying the result for $C'$ in~(\ref{Cpres}) by the factor $\prod_{x_E'} \Delta \tau^{-6}$ previously produced by the integration over $\pi^{ij}$, one finds 
 \be \boxed{C=  \prod_{x_E'} \left[\left(\frac{\Delta V_3 \sqrt{g}}{\pi\hbar\Delta\tau^3}\right)^3 \frac{2\sqrt{3\beta\alpha^5}}{g^2}\right].}  \label{C} \ee
 Like for the scalar field, we think this factor as part of a full integration measure,
 \be \boxed{C \delta g = C \prod_{x_E} \prod_{i\geq j} dg_{ij} (x_E).} \label{IntMeas}\ee 
Eqs.~(\ref{SEQG}),~(\ref{PIEucQG}),~(\ref{C}) and~(\ref{IntMeas}) provide us with the basic formula for the Euclidean path integral in quadratic gravity.  
 
 The integration measure in~(\ref{IntMeas}) is invariant under three-dimensional general coordinate transformations. Let us see why. Under such transformations $g'_{ij}(x') = {\cal J}^k_{~i}{\cal J}^l_{~j} g_{kl}(x)$, where ${\cal J}^k_{~i} \equiv \frac{\partial x^k}{\partial x'^{i}}$, therefore,  at each spacetime point
 \be \frac{\partial g'_{ij}(x')}{\partial g_{kl}(x)} = \frac12 ( {\cal J}^k_{~i}{\cal J}^l_{~j}+{\cal J}^l_{~i}{\cal J}^k_{~j}). \label{gpgchange}\ee 
 By using the methods of Appendix~\ref{Gaussian integral over gravitational momenta} (see the discussion around Eqs.~(\ref{gV}),~(\ref{ggV}) and~(\ref{detgV})), one finds that this Jacobian ``matrix" produces in the path integral a Jacobian determinant $(\det {\cal J})^4$, where ${\cal J}$ is the matrix with elements ${\cal J}^k_{~i}$. On the other hand, $g^2$ in the denominator of~(\ref{C}) produces the same factor under the same transformations, so  the full measure is invariant\footnote{The factor $\Delta V_3 \sqrt{g}$ appearing in~(\ref{C}) is an invariant volume element.}.  
 The presence of an invariant final measure over metrics is a consequence of the fact that we started from the product of the measures associated with canonical coordinates and conjugate momenta, which is invariant.  
 Clearly, the Euclidean action in~(\ref{SEQG}) is also invariant. On the other hand,  The invariance of the probabilities $dq_f \rho(q_f)$, with $\rho(q_f)$ given in~(\ref{probDens}), tells how the states should transform.

The convergence of the Euclidean path integral in~(\ref{PIEucQG}) requires the following conditions in quadratic gravity
\be \alpha>0, \quad \beta>0, \quad \lambda > \frac{\gamma^2}{4\beta},  \label{ConvCondQG} \ee 
which can be derived as follows. First, consider metrics with vanishing $R$.  These metrics can  have a non-vanishing $W^2$.
The integration domain of the path integral include  metric values for which $\sqrt{g}\, W^2$
can be made arbitrarily large keeping $R=0$,
 so this implies $\alpha>0$. Next, consider conformally-flat metrics for which $W^2 = 0$. In this case $R$, together with $g$, can be made arbitrarily large for some field configurations in the domain of path integration
 so $\beta>0$. Finally, the last condition in~(\ref{ConvCondQG}) can be obtained by minimizing the Euclidean Lagrangian with respect to $R$ and requiring $S_E$ to be always positive. Indeed, if such minimum were negative it would be possible to find field configurations with arbitrarily negative $S_E$.
 The conditions in~(\ref{ConvCondQG}) are also sufficient conditions for the convergence of the path integration if, of course, a spacetime discretization is performed. This is opposed to the case of Einstein gravity, where there are no values of the parameters for which the Euclidean path integral could converge as the Euclidean action is always unbounded from below  (the conformal-factor problem). When the continuum limit is taken by removing the cutoff, $\alpha$, $\beta$, $\gamma$ and $\lambda$ generically diverge as a function of the cutoff to ensure the convergence of the physical quantities. In this sense they are bare rather than renormalized quantities, just like $c_4$ in the scalar-field example. For this reason it is not possible to infer  only from~(\ref{ConvCondQG}) the same conditions for the renormalized quantities. For example, one could not claim only from~(\ref{ConvCondQG}) that the theory predicts a positive cosmological constant. 
 
 Note that the discussion on the Hermiticity of $H$  
 that we made for the scalar-field theory (around Eq.~(\ref{ChangeT})) can be made here as well, with the same conclusion, $H^\dagger = H$.

\subsection{Transition to generic coordinate systems} \label{Transition to generic coordinate systems}

So far we have discussed quantum quadratic gravity and constructed the corresponding path integral in the Gauss coordinate system, where the time components of the metric are set to $g_{00}=-1$ and $g_{0i}=0$. Correspondingly, in Euclidean signature, $g_{i4}=0$ and $g_{44}=1$, where we label with 4 the imaginary time. Let us now show how to describe quantum  quadratic gravity in a generic coordinate system. 

This can be done with the help of the Euclidean path integral. To this purpose let us 
insert 
\be 1= \int \delta f \, \delta(f) \ee
with 
\be f_i\equiv g_{i4}, f_4\equiv g_{44}-1, \label{StandGF}\ee
\be \delta f \equiv \prod_{x_E} \prod_\mu df_\mu(x_E), \qquad \delta(f) \equiv \prod_{x_E} \prod_\mu \delta(f_\mu(x_E)) \ee 
as an extra factor in the integrand of~(\ref{PIEucQG}), to obtain
\be \langle q_{f\eta},\tau_f|q_i,\tau_i\rangle^J  = \int^{q(\tau_f) = q_f}_{q(\tau_i) = q_i}  \, C\delta g \,\delta f\, \delta(f) \exp\left(-S_E/\hbar  + \int_{\tau_i}^{\tau_f} d\tau\int d^3x J^{\mu\nu}g_{\mu\nu}\right). \label{PIEucQGf}\ee
where now $S_E$ is computed in the full four-dimensional Euclidean metric. Also, we have taken advantage of the presence of $\delta(f)$ to substitute $J^{ij}g_{ij}$ with $J^{\mu\nu}g_{\mu\nu}$ where $J^{\mu\nu}$ is an arbitrary external  ``current". Note that, as specified before, the product over the imaginary time inside the product over $x_E$ ranges over $\tau \in(\tau_i,\tau_f)$. Therefore, the boundary values of $f_\mu$ at $\tau_i$ and $\tau_f$ are arbitrary.  As a result, the boundary conditions in~(\ref{gKbound}) can be arbitrarily extended by adding any boundary conditions for $f_\mu$, namely for $g_{i4}$ and $g_{44}$. This reflects the freedom of choosing an arbitrary coordinate system (a gauge freedom).

We can express~(\ref{PIEucQGf}) in a more familiar form by considering the change of coordinates, $\xi^\mu \equiv x'^\mu-x^\mu$, that connects  the Gauss coordinate system to another coordinate system with non-vanishing values of $f_\mu$. These $f$s are functionals of the $\xi$s so we can write
\be \langle q_{f\eta},\tau_f|q_i,\tau_i\rangle^J  = \int^{q(\tau_f) = q_f}_{q(\tau_i) = q_i}  \, {\cal D}g \,  \left|\det\frac{\partial f}{\partial\xi}\right|\, \delta(f) \exp\left(-S_E/\hbar  + \int_{\tau_i}^{\tau_f} d\tau\int d^3x J^{\mu\nu}g_{\mu\nu}\right), \label{PIEucQGFP}\ee
where 
\be {\cal D}g \equiv C\delta g \delta\xi, \qquad  \delta \xi \equiv \prod_{x_E} \prod_\mu d\xi^\mu(x_E),  \ee 
and we had to insert the Jacobian $\left|\det\frac{\partial f}{\partial\xi}\right|$ for the change of integration variables from the $f$s to the $\xi$s. This precisely reproduces the Faddeev-Popov (FP) determinant, as we will discuss in Sec.~\ref{EPTTpp} in dealing with the path integral for Green's functions.

As we have seen around Eq.~(\ref{gpgchange}), $C\delta g$ is invariant under three-dimensional general coordinate transformations. The extra factor $\delta \xi$ inside ${\cal D}g$ is invariant under generic change of coordinates that vary the values of $g_{i4}$ and $g_{44}$: this is because one can go to a further coordinate system by shifting $\xi^\mu\to \xi^\mu+\xi'^\mu$, where $\xi'^\mu$ are other spacetime functions, and shifts have unit Jacobians.
 Therefore, the measure ${\cal D}g$ is invariant under four-dimensional general coordinate transformations.

As well-known (see e.g.~Textbook~\cite{Weinberg:1996kr}), matrix elements of the form
\be \int^{q(\tau_f) = q_f}_{q(\tau_i) = q_i}  \, {\cal D}g \,  \left|\det\frac{\partial f}{\partial\xi}\right|\, \delta(f) {\cal O}(x_1) {\cal O}(x_2) ...\exp\left(-S_E/\hbar\right), \label{PIEucQGFPme}\ee
where  ${\cal O}(x_1), {\cal O}(x_2), ...$  represent gauge invariant operators, are independent of the $f$s.  
Then, these  $f$s  clearly play the role of a gauge-fixing function and their choice corresponds to the choice of the coordinate system.

 \subsection{Euclidean path integral for Green's functions} \label{EPTTpp}

 In ordinary theories, the path integral for the matrix elements in~(\ref{PIEuclidean}) can be used, among other things, to construct the path integral for Green's functions. This is also true in the presence of DP variables, at least when an ordinary two-derivative theory is recovered in the large-distance limit. The purpose of this section is to show the latter statement. Before discussing quadratic gravity it is convenient to consider a more general framework as was done at the beginning of Sec.~\ref{EPTTp}. 
 
 A generic Green's function reads 
\be  \langle 0|\mbox{T}\hat q_{i_1}(\tau_1)\hat q_{i_2}(\tau_2)... |0\rangle \ee
where $|0\rangle$ is the eigenstate of $H$ with the minimal real eigenvalue $E_0$, thus $H|0\rangle = E_0 |0\rangle$ (the vacuum). One can express a Green's function in terms of the matrix elements in~(\ref{PIEuclidean}) by inserting the completeness relations
\be \int dq_i |q_i,\tau_i\rangle\langle q_{i\eta},\tau_i|= 1, \qquad \int dq_f |q_f,\tau_f\rangle\langle q_{f\eta},\tau_f|= 1 \ee  
 as follows:
 \be  \langle 0|\mbox{T}\hat q_{i_1}(\tau_1)\hat q_{i_2}(\tau_2)... |0\rangle = \int dq_i dq_f~\Phi_0^*(q_f,-\tau_f)~\langle q_{f\eta},\tau_f|\mbox{T}\hat q_{i_1}(\tau_1)\hat q_{i_2}(\tau_2)...|q_i,\tau_i\rangle~\Phi_0(q_{i\eta},\tau_i),  \label{GreenPhiMat} \ee
 where, for generic $q$ and $\tau$, we have defined
 \be  \Phi_0(q,\tau) \equiv \langle q,\tau|0\rangle  = \exp(-\tau E_0/\hbar) \langle q| 0\rangle \equiv \exp(-\tau E_0/\hbar) \psi_0(q)   \ee
 and we have used $H=H^\dagger$, which we have shown in Sec.~\ref{EPTTp}.
 The quantity $\psi_0(q_\eta)$ defined above is the ground-state wave function. By applying Eq.~(\ref{ProbNorm}) with $A=q$ and $|\psi\rangle=|0\rangle$, one finds that $\psi_0$ allows us to compute the probability density $\rho_0(q)$  of the configuration with canonical coordinates $q$ (in the ground state):
 \be \rho_0(q) = \frac{|\psi_0(q_\eta)|^2}{\int dq'\,|\psi_0(q_\eta')|^2}. \ee  
 Again we can easily check that the unitarity conditions, $\rho_0(q) \geq0$, $\int dq\,\rho_0(q)=1$, hold.
 
 As we have seen, the matrix elements of the T-product in the integrand in~(\ref{GreenPhiMat}) can be obtained through functional derivatives with respect to $J$ of the generating functional in~(\ref{PIEuclideanG}). We then first focus on that functional. Let us choose the current $J$ to be vanishing for $\tau$ outside the interval $[\tau_i,\tau_f]$. Taking two imaginary times $\tau_1$ and $\tau_2$ such that $\tau_1 <\tau_i<\tau_f<\tau_2$, therefore, leads to
 \be \langle q_{2\eta},\tau_2|q_1,\tau_1\rangle^J = \int dq_idq_f~\langle q_{2\eta},\tau_2|q_f,\tau_f\rangle~\langle q_{f\eta},\tau_f|q_i,\tau_i\rangle^J~\langle q_{i\eta},\tau_i|q_1,\tau_1\rangle \label{q2q1J}.\ee 

To  evaluate $\langle q_{i\eta},\tau_i|q_1,\tau_1\rangle$ and $\langle q_{2\eta},\tau_2|q_f,\tau_f\rangle$ in the expression above we first discretize the canonical coordinates, $q$, in a way that their range of variability becomes finite. At this point one would like to compute the matrix $F(M)$ with elements
\be  \left(F(M)\right)_{q_f, q_i} \equiv  \langle q_{f\eta} | e^{-\tau H/\hbar} |q_i\rangle \label{FMdef} \ee 
for any $q_i$, $q_f$ and $\tau$, where $M$ is the matrix with elements 
\be M_{q_f,q_i} \equiv \langle q_{f\eta} |H |q_i\rangle, \label{Mdef} \ee
such that $F(M)=\exp(-\tau M/\hbar)$. 
We are unable to show with present techniques (see~\cite{Bender:2009mq,Bender:2023cem}) that $M$ (and thus $H$) is diagonalizable so we make use of a general decomposition, known sometimes as the Riesz-Dunford formula: 
\be F(M) = \sum_n \left(e^{-\tau \lambda_n/\hbar} {\cal E}^{(0)}_{n} + \sum_{l=1}^{r_n-1} \frac{F^{(l)}(\lambda_n)}{l!} {\cal E}^{(l)}_{n} \right), \label{Riesz-Dunford}\ee
where  the $\lambda_n$ are the eigenvalues of $M$, the $r_n$ are their algebraic multiplicities, in our case $F^{(l)}(\lambda_n) = (-\tau/\hbar)^l\exp(-\tau \lambda_n/\hbar)$  and
\be {\cal E}^{(l)}_{n} \equiv \frac1{2\pi i} \oint_{\Gamma_n} dz~(z-M)^{-1}(z-\lambda_n)^l,  \ee 
with $\Gamma_n$ being a closed curve in the complex plane containing only the eigenvalue $\lambda_n$.  
Note that, knowing $F(M)$, for example through path-integral methods, we can extract the eigenvalue $\lambda_0$ with the smallest real part as follows:
\be \lambda_0 = -\hbar\lim_{\tau \to \infty} \frac1{\tau} \log \left(F(M)\right)_{q_f, q_i} \ee
for arbitrary values of $q_f$ and $q_i$. When the above-mentioned discretization of the $q$s is removed the lowest eigenvalue $\lambda_0$ tends to $E_0$. Note that $E_0$ is real whenever  $\langle q_{f\eta} | e^{-\tau H/\hbar} |q_i\rangle$ is real and positive, which is the case both in the scalar-field example with Lagrangian~(\ref{ScalLag}) and in quadratic gravity, as we have seen in Sec.~\ref{EPTTp}. Moreover, in quadratic gravity it is possible to have $E_0$ non negative  by choosing the bare cosmological constant $\lambda$ large enough, compatibly with the convergence conditions in~(\ref{ConvCondQG}). We will choose $\lambda$ in this way in Sec.~\ref{Quantum cosmology} to ensure that the vacuum energy is positive in agreement with observations. Analogously, in the scalar-field example with Lagrangian in~(\ref{ScalLag}), it is  possible to have $E_0$ non negative  by choosing $V(0)$ large enough. 

In both quadratic gravity and the scalar-field example the low-energy (large-distance) limit is an ordinary two-derivative theory. This can be understood by noting that the eigenvalues of $M$ with the smaller real part  are those for which the effect of the $W^2$ term (or the higher derivative term proportional to $c_4$ in the scalar-field case) is minimized: this occurs when the corresponding state is mostly a combination of nearly four-dimensional homogeneous eigenstates of the canonical coordinates\footnote{In Sec.~\ref{Quantum cosmology} we will provide a more general discussion on the link between the size of an eigenvalue of the Hamiltonian and the configurations that minimize the action and maximize the (absolute value) of the corresponding wave function.}.  In the case of quadratic gravity this ordinary theory is GR plus the scalaron field (the effective scalar associated with the $R^2$ term). Therefore, in the large-distance limit the spectrum of $M$ is an ordinary one.

As a result, $r_0=1$ and in the $\tau_1\to-\infty$ and $\tau_2\to+\infty$ limits, respectively,
\be \langle q_{i\eta},\tau_i|q_1,\tau_1\rangle = e^{-E_0(\tau_i-\tau_1)/\hbar} \psi_0(q_{i\eta})\psi_0^*(q_1), \qquad \langle q_{2\eta},\tau_2|q_f,\tau_f\rangle = e^{-E_0(\tau_2-\tau_f)/\hbar} \psi_0(q_{2\eta})\psi_0^*(q_f). \ee
By inserting these results in~(\ref{q2q1J}) one obtains in the $\tau_1\to-\infty$ and $\tau_2\to+\infty$ limits
\be \frac{e^{E_0(\tau_2-\tau_1)/\hbar}}{\psi_0^*(q_1) \psi_0(q_{2\eta})} \langle q_{2\eta},\tau_2|q_1,\tau_1\rangle^J =  \int dq_idq_f~\Phi_0^*(q_f,-\tau_f)~\langle q_{f\eta},\tau_f|q_i,\tau_i\rangle^J~\Phi_0(q_{i\eta},\tau_i).\ee
On the other hand, for $J=0$, but in the same limit
\be \langle q_{2\eta},\tau_2|q_1,\tau_1\rangle = e^{-E_0(\tau_2-\tau_1)/\hbar} \psi_0(q_{2\eta}) \psi_0^*(q_1).   \ee
Therefore, by comparing with~(\ref{GreenPhiMat})
one obtains that the generating functional of Green's functions, $Z(J)$ can be expressed as 
\be Z(J) =\lim_{\tau_{1,2}\to\mp\infty} \frac{\langle q_{2\eta},\tau_2|q_1,\tau_1\rangle^J}{\langle q_{2\eta},\tau_2|q_1,\tau_1\rangle}, \ee
for arbitrary values of the boundary conditions $q_1$ and $q_2$.

In the case of quadratic gravity, using~(\ref{PIEucQGFP}), 
\be Z(J) = \frac1{``J\to 0"}\int  \, {\cal D}g \,  \left(\det\frac{\delta f}{\delta\xi}\right)\, \delta(f) \exp\left(-S_E/\hbar  + \int d^4x_E \,J^{\mu\nu}g_{\mu\nu}\right), \label{PIEucGF2}\ee
where the prefactor indicates that one has to divide by the same quantity with $J^{\mu\nu}=0$ and we have understood the boundary conditions in the infinite past and future as they are arbitrary. Moreover, we have removed the absolute value around $\det\frac{\partial f}{\partial\xi}$.  Indeed, $\left|\det\frac{\partial f}{\partial\xi}\right|$ is a Jacobian for a change of coordinate and consistency implies $\det\frac{\partial f}{\partial\xi}\neq 0$. So the continuity of $\det\frac{\partial f}{\partial\xi}$ requires that its sign remains constant over the full integration domain. The overall sign can be dropped in~(\ref{PIEucGF2})  because it cancels with the denominator, $``J\to 0"$. Similarly, we have  replaced $\frac{\partial f}{\partial\xi}$ with $\frac{\delta f}{\delta\xi}$ because standard and functional derivatives differ by a factor of the four-dimensional volume element, which again cancels with the denominator, $``J\to 0"$. We recognise $\det\frac{\delta f}{\delta\xi}$ as the FP determinant in the generating functional of Green's functions.

Note that there appear to be no obstacle in going from~(\ref{PIEucGF2}) to the generating functional of the Lorentzian Green's functions. One should replace the Euclidean  ``current" $J^{\mu\nu}(\tau)$ with a Lorentzian current with the substitutions $J^{kl}(\tau)\to J^{kl}(it)/\hbar$, $J^{k4}(\tau)\to iJ^{k0}(it)/\hbar$ and $J^{44}(\tau)\to -J^{00}(it)/\hbar$, where we have understood the dependence on the spatial coordinates. At the same time we should perform a change of variables in path integral of the form $g_{kl}(\tau)\to g_{kl}(it)$, $g_{k4}(\tau)\to -ig_{k0}(it)$ and $g_{44}(\tau)\to -g_{00}(it)$. As already noted, the boundary conditions in~(\ref{PIEucGF2}) are arbitrary so one can choose some that are compatible with a real Lorentzian metric with the chosen signature: e.g. $g'_{ij}=0$, $g_{i4}=0$, $g_{44}=1$ both at initial and final times. So these substitutions lead to
\be {\cal Z}(J) = \frac1{``J\to 0"}\int  \, {\cal D}g \,  \left(\det\frac{\delta f}{\delta\xi}\right)\, \delta(f) \exp\left(iS/\hbar  + i\int d^4x \,J^{\mu\nu}g_{\mu\nu}/\hbar\right), \label{PIEucGF2L}\ee
where ${\cal Z}(J)$ is the generating functional of the Lorentzian Green's functions.
It should, however, be kept in mind that, as always, the Lorentzian path integral is a formal object, precisely defined only as an analytic continuation of the Euclidean one.

At this point one would be interested in showing that previous perturbative calculations performed expanding around the Minkowski metric $\eta_{\mu\nu}$ are recovered when applicable.  Since the FP determinant multiplies $\delta(f)$, one only needs to compute it for vanishing $\xi^\mu$. So its $\xi$ dependence can be dropped inside the integral in~(\ref{PIEucGF2L}). Moreover, $S$ is also $\xi$-independent as invariant under general coordinate transformations. So, when one considers the path integral for the vacuum expectation value (VEV) of T-products of gauge invariant operators, the only integrand factor that depends on $\xi^\mu$ is $\delta(f)$. For the simple $f$ chosen in~\cite{Stelle:1976gc} (and other quantum calculations around the Minkowski background, e.g.~\cite{Salvio:2018kwh}) the only role of this $\delta$ function is to eliminate the zero modes of the differential operator appearing in the linearized equations for $h_{\mu\nu}\equiv g_{\mu\nu}-\eta_{\mu\nu}$. Furthermore, note that the local factors of $g$ in the integration measure (see~(\ref{C})) can be dropped in doing perturbative calculations around the Minkowski background based on dimensional regularization, as explained in~\cite{Stelle:1976gc}. So, modulo gauge artefacts that do not affect physics, the path integral used in previous perturbative calculations around $\eta_{\mu\nu}$ is reproduced.

 \section{The quadratic gravity effective action}\label{The quadratic gravity effective action}

The ``current" $J_{\mu\nu}$ in~(\ref{PIEucGF2L}) can also be interpreted as an external physical source for the system, corresponding to a sort of energy-momentum tensor. Such a tensor can be produced by a set of matter fields coupled to quadratic gravity. Indeed, in this physical situation one should add to the starting action of quadratic gravity in~(\ref{TotAction}) a matter term. Repeating the same steps that we have performed in the absence of matter fields one obtains a generating functional of the form~(\ref{PIEucGF2L}), where, however, the Green's functions are to be obtained by performing functional derivatives with respect to $J^{\mu\nu}$ and then setting $J^{\mu\nu}$ to the physical energy-momentum tensor in question.

  In this section we extend the concept of (quantum) effective action, which is typically derived for scalar fields in textbooks, to the case of the non-perturbative and background-independent formulation of quadratic gravity we have given\footnote{For a description of the effective action in previous approaches to quantum gravity see~\cite{Buchbinder:1992rb}.}. 

With the described physical situation in mind 
\be \langle g_{\mu\nu}\rangle_J \equiv \frac1{``J\to 0"}\frac{\int  \, {\cal D}g \,  \left(\det\frac{\delta f}{\delta\xi}\right)\, \delta(f) \, g_{\mu\nu} \, \exp\left(iS/\hbar  + i\int d^4x \,J^{\mu\nu}g_{\mu\nu}/\hbar\right)}{\int  \, {\cal D}g \,  \left(\det\frac{\delta f}{\delta\xi}\right)\, \delta(f) \exp\left(iS/\hbar  + i\int d^4x \,J^{\mu\nu}g_{\mu\nu}/\hbar\right)}\ee
can be interpreted as the expectation value of $g_{\mu\nu}$ in the presence of the physical energy-momentum tensor, which we still call $J^{\mu\nu}$ here for simplicity. Note that $\langle g_{\mu\nu}\rangle_J$ is real because it is the expectation value of a canonical variable~\cite{Salvio:2020axm}. Defining the new functional $W(J)$ through ${\cal Z}(J)\equiv \exp(i W(J)/\hbar)$ one easily finds 
\be \frac{\delta W}{\delta J^{\mu\nu}} = \langle g_{\mu\nu}\rangle_J. \ee

The quadratic gravity effective action $\Gamma$ can then be defined as a funcional of $\langle g_{\mu\nu}\rangle_J$ by 
\be \Gamma(\langle g_{\mu\nu}\rangle_J)\equiv W(J)- \int d^4x J^{\mu\nu}\langle g_{\mu\nu}\rangle_J, \label{GammaDef}\ee 
where $J$ has to be thought of as a functional of $\langle g_{\mu\nu}\rangle_J$.
From this definition it follows that $\Gamma$ satisfies 
\be \frac{\delta\Gamma}{\delta \langle g_{\mu\nu}\rangle_J} + J^{\mu\nu} = 0.  \ee
In words, $\Gamma$ is an action functional of $\langle g_{\mu\nu}\rangle_J$, whose field equations allow to determine $\langle g_{\mu\nu}\rangle_J$. This is the reason why $\Gamma$ is called the quadratic gravity effective action.

Therefore, this effective action, used at tree level, captures the full quantum effects. In Sec.~\ref{The classical limit} a semi-classical expansion of $\Gamma$ will be performed.

\section{The classical limit}\label{The classical limit}

One of the advantages of the path integral formula is the fact that it allows us to take the classical limit, $\hbar \to 0$, easily. Taking this limit in~(\ref{PIEucvarphi}) or~(\ref{PIEucQGf}) (for the four-derivative scalar-field theory and quadratic gravity, respectively) and setting the external currents to zero one obtains
\be \langle q_{f\eta},\tau_f|q_i,\tau_i\rangle = \exp(-S_E^{\rm ren}/\hbar) \label{classicalAmp}\ee
where $S_E^{\rm ren}$ is the classical Euclidean action with renormalized coefficients and computed in the solution of the  imaginary-time classical equations of motion satisfying the boundary conditions $q(\tau_i) = q_i$ and $q(\tau_f) = q_f$. These are generic initial and final conditions for $g_{ij}$ and $g'_{ij}$ (or $\varphi$ and $\varphi'$ in the scalar-field example). In other words all possible imaginary-time classical motions (for arbitrary boundary conditions) are described by the imaginary-time classical theory associated with $S_E^{\rm ren}$. This action can be used to compute instantons, see e.g.~\cite{Salvio:2016mvj}. Now, focusing on quadratic gravity and going back to real time one obtains that the classical theory is described by our starting action in~(\ref{TotAction}), but with renormalized coefficients\footnote{In the notation used in Refs.~\cite{Salvio:2014soa,Salvio:2017qkx,Salvio:2018kwh,Salvio:2019wcp} by the same author, we have $\alpha_r=1/f_2^2$, $\beta_r=-1/(6f_0^2)$, $\gamma_r=-\bp^2/2$ and $\lambda_r=\Lambda$, where $\bp$ is the reduced Planck scale and $\Lambda$ is the cosmological constant. One-loop renormalization group equations (RGEs) have been obtained in the modified minimal subtraction scheme in Refs.~\cite{Avramidi:1985ki} and later extended to a generic renormalizable matter sector in~\cite{Salvio:2014soa,Salvio:2017qkx}. The RGEs generally depend on the renormalization scheme even at one-loop level~\cite{Buccio:2024hys}.} $\alpha_r$, $\beta_r$, $\gamma_r$ and $\lambda_r$, instead of the bare ones, $\alpha$, $\beta$, $\gamma$ and $\lambda$, respectively (we call this action $S^{\rm ren}$). Then we see  that, although the extra canonical coordinates ($K_{ij} = -\dot g_{ij}/2$, or $\varphi_2 = \dot\varphi$) have a purely imaginary spectrum the corresponding classical observable quantities remain real.

This result can also be independently obtained by performing a semi-classical expansion of  the (quantum) effective action. Since this expansion is well known for scalar fields we directly focus on quadratic gravity and make use of the effective action defined in Sec.~\ref{The quadratic gravity effective action}.  

In the classical limit, $\hbar\to0$, the path integral 
in~(\ref{PIEucGF2L}) is dominated by a metric satisfying 
\be \frac{\delta S}{\delta g_{\mu\nu}} +J^{\mu\nu} = 0. \ee
Let us  call such a metric $g^J_{\mu\nu}$ to highlight its dependence on $J_{\mu\nu}$. It then makes sense to perform a change of variables $g_{\mu\nu} = g^J_{\mu\nu} + h_{\mu\nu}$ in~(\ref{PIEucGF2L}) and expand the integrand in powers of the fluctuation field $h_{\mu\nu}$. This semi-classical perturbative approach allows us, among other things, to reconstruct previous perturbative quantizations of quadratic gravity performed starting from a given classical background, see e.g.~\cite{Deruelle:2010kf,Kannike:2015apa,Salvio:2017xul}. Doing so one obtains
\bea {\cal Z}(J) &=& \frac1{``J\to 0"}\int  \, {\cal D}_J h \,  \left(\det\frac{\delta f}{\delta\xi}\right)\, \delta(f) \nonumber \\ 
&&\exp\left(\frac{i}{\hbar}S|_{g^J}  + \frac{i}{\hbar}\int d^4x \,J^{\mu\nu}g^J_{\mu\nu} + \left.\frac{i}{2\hbar} \frac{\delta^2 S}{\delta g_{\mu\nu} \delta g_{\rho\sigma}}\right|_{g^J}h_{\mu\nu}h_{\rho\sigma} + ...\right), \label{SC-exp}\eea
where the symbol $...|_{g^J}$ means that we are evaluating the corresponding quantity ...  at $g_{\mu\nu}=g^J_{\mu\nu}$ and ${\cal D}_Jh$ 
is the measure ${\cal D} g$ expressed in terms of $h_{\mu\nu}$. The only property of this measure that we will use is its independence\footnote{The $\hbar$ in~(\ref{C}) cancels with the denominator, $``J\to 0"$.} of $\hbar$. The first term in this expansion for $\hbar$ small gives
\be {\cal Z}(J) = \exp\left(\frac{i}{\hbar}S|^{\rm ren}_{g^J}-\frac{i}{\hbar}S|^{\rm ren}_{g^0}  + \frac{i}{\hbar}\int d^4x \,J^{\mu\nu}g^J_{\mu\nu}\right), \qquad (\hbar\to0) \ee
where $S|^{\rm ren}_{g^J}$ is the starting action~(\ref{TotAction}) with renormalized coefficients  and computed in the corresponding $g^J_{\mu\nu}$, namely the solution of
\be \frac{\delta S^{\rm ren}}{\delta g_{\mu\nu}} +J^{\mu\nu} = 0. \ee
 The equation above can be  equivalently written as 
\be W(J) = S^{\rm ren}|_{g^J}-S^{\rm ren}|_{g^0}  + \int d^4x \,J^{\mu\nu}g^J_{\mu\nu}, \qquad (\hbar\to0). \ee 
So, in the classical limit one obtains $g^J_{\mu\nu} = \langle g_{\mu\nu}\rangle_J$ and, 
using~(\ref{GammaDef}), 
\be   \Gamma(\langle g_{\mu\nu}\rangle_J) = S|^{\rm ren}_{g^J}-S|^{\rm ren}_{g^0}, \qquad (\hbar\to0) \ee 
and thus 
\be 0=\frac{\delta\Gamma}{\delta \langle g_{\mu\nu}\rangle_J} + J^{\mu\nu}=\frac{\delta S|^{\rm ren}_{g^J}}{\delta g_{\mu\nu}^J} + J^{\mu\nu}.  \ee
Again one finds that the classical limit is described by the classical action in~(\ref{TotAction}) with renormalized coefficients. This action has been used to find several classical solutions, e.g.~describing astrophysical objects of interest such as black holes~\cite{Lu:2015cqa,Lu:2015psa,Lu:2017kzi,dePaulaNetto:2023vtg} or horizonless ultracompact objects\footnote{Since in quadratic gravity the strength of gravity, unlike in Einstein's GR, decreases below a critical (microscopic) length scale~\cite{Giudice:2014tma,Salvio:2016vxi,Salvio:2020axm}, the gravitational collapse stops once the matter distribution is compressed below that scale. So solutions with extended matter distributions, such as those in~\cite{Holdom:2016nek,Holdom:2019bdv,Salvio:2019llz,Aydemir:2020xfd,dePaulaNetto:2023vtg}, appears to be particularly relevant.}~\cite{Holdom:2016nek,Holdom:2019bdv,Salvio:2019llz,Aydemir:2020xfd}. The result obtained in this section supports these findings.

One might wonder how a quantum theory with a bounded spectrum of the Hamiltonian operator could lead in the classical limit to a theory which, as we have seen, has a classical Hamiltonian that is unbounded from below (although in the case of classical quadratic gravity the runaway rates can be slow enough to ensure compatibility with observations).

Note that in ordinary theories the non-negativity of the spectrum of the Hamiltonian operator implies that the classical Hamiltonian $H_c$ is also non negative. This can be seen from the imaginary-time Hamiltonian path integral for zero external current
\be \langle q_{f},\tau_f|q_i,\tau_i\rangle \equiv \int^{q(\tau_f) = q_f}_{q(\tau_i) = q_i} \delta q \delta p~\exp\left( \frac1{\hbar}\int_{\tau_i}^{\tau_f} d\tau (i pq'-\bar H(q,p))\right)\label{ITHPI}\ee
and its real-time counterpart
\be \langle q_{f},t_f|q_i,t_i\rangle \equiv \int^{q(t_f) = q_f}_{q(t_i) = q_i} \delta q \delta p~\exp\left( \frac{i}{\hbar}\int_{t_i}^{t_f} dt (p\dot q-\bar H(q,p))\right)\label{RTHPI}
\ee
having defined $|q,t\rangle\equiv \exp(i H t/\hbar)|q\rangle$ and $\langle q,t|\equiv \langle q|\exp(-i H t/\hbar)$ for real time $t$. In ordinary theories $\bar H$ is real in the classical limit because it is a real function of ordinary canonical variables with a real spectrum. Using in~(\ref{RTHPI}) the stationary phase approximation, which becomes exact in the $\hbar\to 0$ limit, then allows us to conclude that $\bar H$ coincides with $H_c$. 
Let us now go to the imaginary-time path integral in~(\ref{ITHPI}) and suppose that $\bar H(q,p)$ were negative for some $q$ and $p$, say $q_0$ and $p_0$.  Choosing $q_i=q_0$ and sending $\tau_f\to+\infty$ and $\tau_i\to -\infty$ one would obtain a divergent integral over $dp(\tau_i)$ because the path integral would feature an exponent with a positive real part of the form $\int_{-\infty}^{+\infty}d\tau (-\bar H(q_0,p_0))$ in some finite integration domain of $dp(\tau_i)$. This cannot happen if the spectrum of the Hamiltonian operator is non negative because  in this case, inserting a complete set, $|n\rangle$, of eigenstates of $H$ with eigenvalues $E_n$ one finds
 \be \langle q_{f},\tau_f|q_i,\tau_i\rangle  =  \sum_n e^{-E_n(\tau_f-\tau_i)/\hbar} \langle q_f|n\rangle\langle n|q_i\rangle,  \ee 
 which converges as $\tau_f\to+\infty$ and $\tau_i\to +\infty$ if $E_n\geq 0$.
 
 In theories with DP variables this conclusions cannot be reached when $\bar H$ is not real, like in quadratic gravity or the scalar-field example we have discussed (see e.g.~Eq.~(\ref{barHSF})). The reason is that in that case the real-time Hamiltonian path integral would no longer imply that $\bar H$ is $H_c$ because the path integrand would feature an exponent with both an imaginary and a real part and the stationary phase approximation cannot be used to conclude $\bar H=H_c$. However, when the ground state energy $E_0$ is not negative one can still say that the real part of $\bar H$ is never negative, which is the real universal implication of a non-negative spectrum of $H$. 
 
 In theories with DP variables with a consistent Euclidean path integral, featuring a positive Euclidean action $S_E$, the classical Hamiltonian $H_c$ should be determined by taking the classical limit to find~(\ref{classicalAmp}) and then analytically continue to real time. 
 
 Let us conclude this section by noting that  in a given physical situation one has to check whether the observed value of $\hbar$ is small enough for the classical limit to be a good approximation. It is indeed possible that quantum physics overcomes classical physics in some situations of interest~\cite{Donoghue:2021eto}.

\section{Quantum cosmology}\label{Quantum cosmology}

In this section we apply the theory we have previously constructed  to quantum aspects in cosmology. 

First, let us observe that the perturbative tree-level calculations performed in quadratic gravity expanding around cosmological FLRW metrics~\cite{Deruelle:2010kf,Kannike:2015apa,Salvio:2017xul} are reproduced by our non-perturbative construction. This is because in those calculations it was only assumed that canonical commutators hold and that the spectrum of the Hamiltonian is bounded from below (both these conditions are satisfied here). It is also interesting to note that the fakeon prescription~\cite{Anselmi:2017ygm,Anselmi:2020lpp}, which eliminates the states with negative norm (with respect to the indefinite inner product), has led so far to the same observable predictions when applicable~\cite{Salvio:2022mld,Dondarini:2023atl}. Moreover, note that we can apply the Green's functions we have constructed to study non-time-ordered VEV of cosmological perturbations (which are relevant to obtain power spectra): one can consider the  time-ordered product of two cosmological perturbations computed at times $t+\epsilon$ and $t$ and then let $\epsilon\to 0^+$ (and then generalize this technique to the time-ordered product of more cosmological perturbations in a  straightforward way).
 
 In quantum cosmology, as pointed out by Hartle and Hawking~\cite{Hartle:1983ai}, one describes the universe through a quantum-mechanical wave function (actually a wave functional). The Hartle-Hawking construction was, however, based on Einstein's gravity, whose Euclidean path integral suffers from the conformal-factor problem. As observed in Sec.~\ref{EPTTp}, the Euclidean path integral in the quantization of quadratic gravity we have specified does not suffer from this problem. Here we can, therefore, implement the Hartle-Hawking idea in a more rigorous way. 
 
 For example, the ground-state wave functional $\psi_0(q_{\eta})$, which we have previously introduced in Sec.~\ref{EPTTpp}, is given here by
\be {\cal N}_0 \psi_0(q_{f\eta}) =\lim_{\tau\to +\infty} e^{E_0\tau/\hbar}\langle q_{f\eta} | e^{-\tau H/\hbar} |q_i\rangle , \label{GSwf}\ee
where 
\be \langle q_{f\eta} | e^{-\tau H/\hbar} |q_i\rangle = \int^{q(\tau) = q_f}_{q(0) = q_i}  \, {\cal D}g \,  \left|\det\frac{\partial f}{\partial\xi}\right|\, \delta(f)  \exp\left(-S_E/\hbar\right), \ee 
modulo a $q_f$-independent proportionality factor, ${\cal N}_0 = \psi_0^*(q_i)$, that does not affect physics. This means that the ground-state wave function of the universe can be computed regardless of the past boundary condition. The result obtained here is reminiscent of the no-boundary wave function of the universe~\cite{Hawking:1981gb,Hartle:1983ai} (see also~\cite{Lehners:2023yrj,Maldacena:2024uhs} for recent reviews), which, however, was proposed within Einstein gravity that is affected by the conformal-factor problem, unlike quadratic gravity. The smallest eigenvalue $E_0$  corresponding  to $\psi_0$, namely the vacuum energy, that appears in~(\ref{GSwf}) can be computed using 
\be E_0 =- \hbar\lim_{\tau \to \infty} \frac{1}{\tau} \log \langle q_{f\eta} | e^{-\tau H/\hbar} |q_i\rangle. \ee
This limit is generically divergent and one should rather consider the vacuum energy density $\rho_V$, that is $E_0$ divided by the three-dimensional volume element. It is easy to check, at least in the classical limit when~(\ref{classicalAmp}) holds, that $\rho_V$ vanishes when $\lambda_r=0$, as expected. 
One can reiterate this procedure to reconstruct all the large-distance spectrum. For example, a first excited state has eigenvalue
\be E_1 = - \hbar\lim_{\tau \to \infty} \frac{1}{\tau} \log \left(\langle q_{f\eta} | e^{-\tau H/\hbar} |q_i\rangle - \psi_0(q_{f\eta})\psi_0^*(q_i)e^{-E_0\tau/\hbar} \right). \ee

The wave functionals satisfy a sort of Wheeler-DeWitt equation~\cite{DeWitt:1967yk,Wheeler}. For example, for the ground state this equation in our notation simply reads
\be \langle q_\eta|H|\psi_0\rangle = E_0\langle q_\eta|\psi_0\rangle. \ee
Indeed, derivatives appear on the right-hand-side  when we note that if $\{q,p\}$ is an ordinary pair of canonical variables $\langle q|\hat p=-i\frac{d}{dq}\langle q|$, while if $\{q,p\}$ is a DP pair of canonical variables $\langle q|\hat p=\frac{d}{dq}\langle q|$, as explained in~\cite{Salvio:2015gsi,Salvio:2018crh}. Since we are dealing with a field theory here (namely quadratic gravity), these derivatives are actually functional derivatives.

As discussed in Sec.~\ref{EPTTpp}, in the large-distance limit, the smaller eigenvalues of $H$ are those for which the effect of the $W^2$ term is minimized.
The $W^2$ term vanishes for conformally-flat metrics. This means that such eigenvalues are those whose corresponding wave functionals $\langle q_\eta  |\psi\rangle$ are peaked around values of the canonical coordinates, $q$, of the form 
\be g_{ij}(\vec x) = \Omega^2(\vec x) \delta_{ij},  \qquad K_{ij}(\vec x) = \kappa(\vec x) \delta_{\ij}. \label{ConformalgK} \ee 
(later in this section we will provide a more general explicit discussion about this correspondence).
Note that  
a wave functional $\langle q_\eta  |\psi\rangle$ that is peaked around some metric configurations also corresponds to a probability density $\rho(q)$ that is peaked around the same configurations: using Eq.~(\ref{ProbNorm}) with $A =q$
 \be \hspace{-.65cm}\rho(q) = \frac{|\langle q_{\eta}|\psi\rangle|^2}{\langle \psi| \eta |\psi\rangle} = \frac{|\langle q_{\eta}|\psi\rangle|^2}{\int dq_\eta |\langle q_\eta|\psi\rangle|^2}, \label{probD} \ee
 which, as always, satisfies the unitarity conditions in~(\ref{unitaryrho}).
In the case of quadratic gravity a wave function $\langle q_\eta  |\psi\rangle$ is  a functional of the metric (and its time derivative) at a given time. Therefore, the field configurations in~(\ref{ConformalgK}) depend only on the three-dimensional spatial coordinate $\vec x$. Identifying $K_{ij}$ with $-\dot g_{ij}/2$ in~(\ref{ConformalgK}) gives a Lorentzian conformally-flat metric, which is what we want to make contact with real-time cosmology. In cosmology conformal flatness con only be broken  by inhomogeneities and anisotropies as the FLRW metric is conformally flat. However, we can construct specific conformally-flat metrics that are not  homogeneous and isotropic. Regarding the latter point, note that, in the large-distance limit, the smallest eigenvalues of $H$ also corresponds to nearly constant values of the Ricci scalar $R$: indeed, the Euclidean action in~(\ref{SEQG}) acquires its minimum for metrics that are conformally flat  and have a spacetime-independent Ricci scalar given by
\be \bar R = -\frac{\gamma}{2\beta}. \ee 
 Large inhomogeneities and anisotropies in cosmology necessarily corresponds to large values of (the real part of) the eigenvalues of $H$. 

These results are related to those of Refs.~\cite{Lehners:2019ibe,Lehners:2023fud}, where the finite-action principle was proposed to explain homogeneity and isotropy of the universe within quadratic gravity in the classical limit: in our fully quantum construction the configuration that leads to the smallest action corresponds to peaks in the wave functional of the ground state and are homogeneous and isotropic, while inhomogeneities and anisotropies  lead to a larger action and are associated with excited states. 

The link between the size of an eigenvalue $E$ of $H$ and the configurations that minimize the action and maximize the (absolute value) of the corresponding wave functional $\langle q_\eta|\psi\rangle$ can more generally seen as follows. Note 
\be \int dq_i \langle q_{f\eta},\tau_f|q_i,\tau_i\rangle \langle q_{i\eta}|\psi\rangle = e^{-(\tau_f-\tau_i)E/\hbar}\langle q_{f\eta}|\psi\rangle. \ee
Multiplying by $\langle\psi|q_{f\eta}\rangle$ and integrating over $q_f$ on both sides leads to
\be \int dq_fdq_i \langle\psi|q_{f\eta}\rangle \langle q_{f\eta},\tau_f|q_i,\tau_i\rangle \langle q_{i\eta}|\psi\rangle  = e^{-(\tau_f-\tau_i)E/\hbar}  \int dq_f  |\langle q_{f\eta}|\psi\rangle|^2\ee
The integral $\int dq_f  |\langle q_{f\eta}|\psi\rangle|^2$ can be set to 1 with a conventional\footnote{Indeed, as usual, a change of normalization of $|\psi\rangle$ leaves the probability density in~(\ref{probD}) invariant.} choice of the normalization of $|\psi\rangle$, which leads to 
\be \int dq_fdq_i \langle\psi|q_{f\eta}\rangle \langle q_{f\eta},\tau_f|q_i,\tau_i\rangle \langle q_{i\eta}|\psi\rangle  = e^{-(\tau_f-\tau_i)E/\hbar}.  \ee
Therefore, recalling that in quadratic gravity $\langle q_{f\eta},\tau_f|q_i,\tau_i\rangle$ is given by~(\ref{PIEucQG}) (or for a generic gauge by~(\ref{PIEucQGFP})) with $J^{\mu\nu}=0$, the wave functionals with the smaller (real part of) $E$ are those somewhat peaked around configurations for which the Euclidean action is small.

It is interesting to note that in the small $\alpha$ limit\footnote{One might be worried by the fact that in this limit the path-integration measure in~(\ref{PIEucQG}) is singular because $C$ in~(\ref{C}) contains some power of $\alpha$. However, note that $\alpha$ disappears from the measure when one considers the probability densities because it cancels with an equal factor coming from the denominator in~(\ref{probDens}). Similarly, any field-independent factor in~(\ref{C}) does not contribute to the probability densities. The small $\alpha$ limit has been taken together with the small $\gamma$ and $\lambda$ limit in Refs.~\cite{Edery:2023hxl,Hell:2023mph}. For a discussion of the large $\alpha$ limit  see~\cite{Maldacena:2011mk,Hinterbichler:2015soa,Salvio:2017xul,Hell:2023rbf,Buoninfante:2023ryt}.} the theory reduces to Starobinsky's model of inflation. Therefore, our construction can also be applied to study non-perturbative aspects of this successful inflationary model. However, note that in that limit the $W^2$ term becomes inactive and, therefore, the smallest eigenvalues of $H$ no longer necessarily correspond to wave functionals peaked around conformally-flat metrics.

One important implication of the formulation of quadratic gravity presented here is the fact that  the homogeneity and isotropy of the initial conditions for inflation are explained rather than postulated. Let us divide the space at initial time $t_i$ in  causally disconnected patches. The state vector at time $t_i$ of such a system can be described by a direct product of the form
\be |\psi(t_i)\rangle = |\psi_I\rangle|\psi_{II}\rangle ... |\psi_K\rangle ...  \ee
where the factor $|\psi_K\rangle$ describes the $K$th patch.  Let us focus on a given patch at a time as its time evolution $|\psi_K(t_i)\rangle \to|\psi_K(t)\rangle$ 
is independent of the others, at least for small enough $t-t_i$.  Suppose to start from a highly inhomogeneous and/or anisotropic patch at time $t_i$. This corresponds to the short-distance limit as we have seen. Then we can be in two distinct situations. Either (i) there is a dynamical mechanism that makes the field configuration homogeneous and isotropic at large time or (ii) such a mechanism is not present. In Case (i) the homogeneity and isotropy of the initial condition for inflation are clearly explained. In Case (ii) we can either be in a situation (iia)  where the initial state becomes nearly classical as time passes  or in a complementary situation (iib) where the initial state remains quantum. In Situation (iia) the homogeneity and isotropy of the initial condition for inflation is explained by the classical mechanism described in Ref.~\cite{Salvio:2019ewf}, based on anthropic selection, which has been reviewed in the introduction. In Situation (iib) life remains impossible because classical entities, such as liquid water, are essential for life as we know it; therefore, a similar anthropic exclusion of the considered patch takes place.

\begin{figure}[t]
    \centering
    \includegraphics[width=0.8\textwidth]{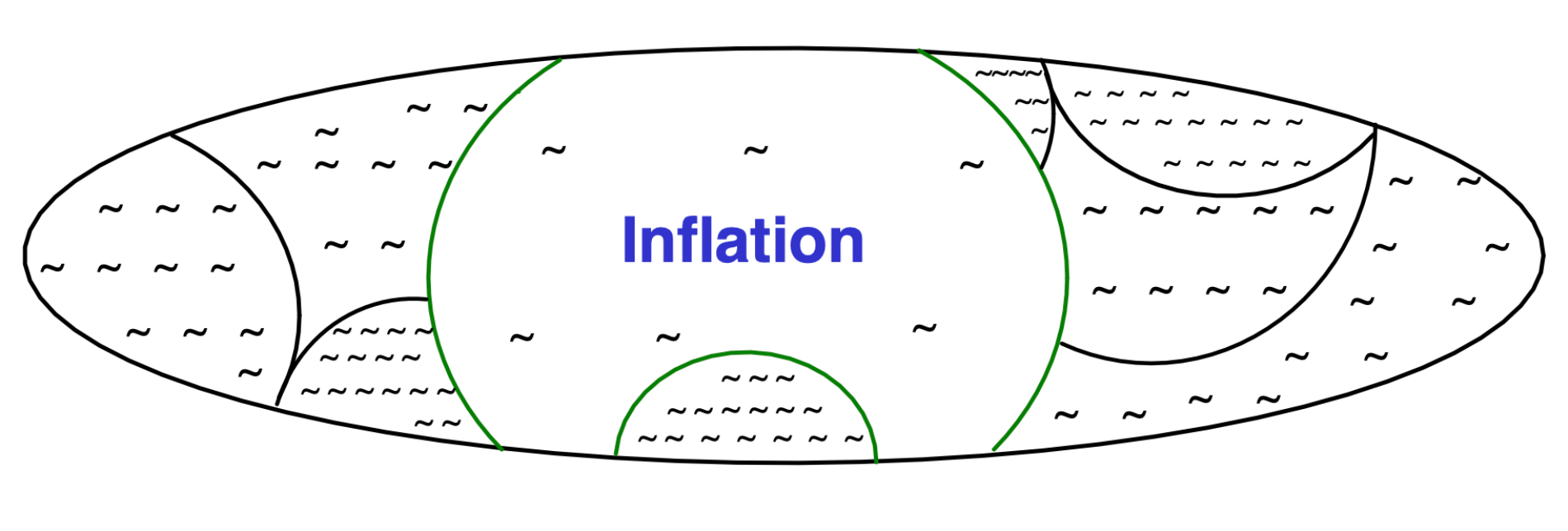}
    \caption{\em Cartoon for the primordial universe. The density of wiggle represents the amount of inhomogeneity and anisotropy.}
    \label{patches}
\end{figure}

It is only in the nearly homogeneous and isotropic  patches that inflation can occur and observations can eventually be made. If such a patch has undergone exponential expansion an  internal point (far from its border) remains causally disconnected from the other patches even at large $t-t_i$. This mechanism is an explicit quantum realization of Linde's chaotic inflation~\cite{Linde:1983gd}. The situation is illustrated in Fig.~\ref{patches}: the nearly homogeneous and isotropic patches are compatible with life and can inflate.

  It is important to note that even if there were non standard features of the spectrum of $H$ (i.e.~$H$ not diagonalizable with only real eigenvalues) there would be no way to observe such features. Indeed, suppose we were in such a situation and let us start from a highly inhomogeneous and anisotropic initial patch, such that the corresponding state $|\psi_K\rangle$ could have a non-negligible overlap with an eigenstate $|\psi_c\rangle$ of $H$ with a complex eigenvalue $E_c=E_R+iE_I$, where $E_R$ and $E_I$ are real numbers (with $E_I$ being non negligible), and/or $|\psi_K\rangle$ cannot be written as a linear combination of eigenstates of $H$. Since the time evolution is given by the operator $\exp(-i(t_f-t_i)H/\hbar)$, a state with $E_I<0$ would decay with time. However, if there were a complex eigenvalue the spectrum would also include its complex conjugate because $H$ is time-reversal invariant\footnote{So, $HT|\psi_c\rangle =TH|\psi_c\rangle=E_c^*T|\psi_c\rangle$ given the antilinearity of $T$ and, thus,  an eigenstate of $H$ with eigenvalue $E_c^*$ is $T|\psi_c\rangle$.}, $T H T^{-1} = H$ . This means that $E_I>0$ would also be realized and we must take this possibility into account. However, in a situation where $|\psi_K\rangle$ had a non-negligible overlap with an eigenstate $|\psi_c\rangle$ with $E_I>0$, Eq.~(\ref{probDens}) would tell us that, as $t_f-t_i$ becomes large, the density probability of patch $K$ exponentially converges  to 
\be \rho_K(q_f) = \frac{|\langle q_{f\eta}|\exp(-i(t_f-t_i) E_R/\hbar+(t_f-t_i) E_I/\hbar)|\psi_c\rangle|^2}{\int dq_\eta |\langle q_\eta|\exp(-i(t_f-t_i) E_R/\hbar+(t_f-t_i) E_I/\hbar)|\psi_c\rangle|^2} = \frac{|\langle q_{f\eta}|\psi_c\rangle|^2}{\int dq_\eta |\langle q_\eta|\psi_c\rangle|^2}, \ee 
where $|\psi_c\rangle$ is the eigenstate with a non negligible overlap with $|\psi_K\rangle$ and with the largest value of $E_I$. 
In words, the probability density of such a patch would converge to that associated with a state $|\psi_c\rangle$ and stay there forever. As we have seen, such a state must be highly inhomogeneous and/or anisotropic and, thus, cannot be compatible with observers. If $|\psi_K\rangle$ could not be  written as a linear combination of eigenstates of $H$ we would obtain a similar result, except that, if there are no complex eigenvalues, the probability density would converge linearly rather than exponentially to 
\be \rho_K(q_f)  = \frac{|\langle q_{f\eta}|\psi_m\rangle|^2}{\int dq_\eta |\langle q_\eta|\psi_m\rangle|^2}, \ee 
where $|\psi_m\rangle$ is now the eigenstate of $H$ with the largest algebraic multiplicity, in the discrete approximation of Eq.~(\ref{Riesz-Dunford}). Again, observations in such a patch cannot be made. Furthermore, such a patch does not inflate because, as we have seen, the inhomogeneities are not diluted as time passes.

\section{\color{verdes}Conclusions}\label{conclusions}

Here is a detailed summary of the results obtained together with conclusions.

\begin{itemize}
\item In the first section (after the introduction) the $3+1$ formalism was introduced for quadratic gravity featuring both the $R^2$ and $W^2$ terms in addition to the standard Einstein-Hilbert and cosmological constant terms. The simple Gauss coordinate system was chosen postponing the restoration of general covariance to a subsequent section, Sec.~\ref{Transition to generic coordinate systems}. All basic tensors were provided in the $3+1$ formalism in terms of the metric, $g_{ij}$, $K_{ij}\equiv -\dot g_{ij}/2$ and $\dot K_{ij}$ in Sec.~\ref{The action} and Appendix~\ref{Curvature tensors}.
\item In Sec.~\ref{Canonical quadratic gravity} Ostrogradsky's canonical method was non-perturbatively applied to quadratic gravity to obtain the canonical coordinates, $g_{ij}$ and $K_{ij}$, and the corresponding conjugate momenta, $\pi^{ij}$ and $P^{ij}$. When both the $R^2$ and $W^2$ terms are present, $\dot K_{ij}$ can be expressed in terms of $g_{ij}$, $K_{ij}$ and $P^{ij}$. Then, the {\it classical} Hamiltonian $H_c$  can be expressed in terms of the canonical and $H_c$ depends on $\pi^{ij}$ only through the term in~(\ref{UnbTerm}). As a result, $H_c$ is unbounded from below (and above). In previous works~\cite{Salvio:2019ewf,Gross:2020tph}, however, it was shown that the runaway rates in classical quadratic gravity can be slow enough to ensure agreement with observations (in this sense one has metastability). This occurs when the typical length scales $L$ of inhomogeneities and anisotropies are much above a threshold  that is low enough to describe the whole cosmology~\cite{Salvio:2019ewf}.  This threshold is $L\gg L_2\equiv\sqrt{2/f_2\bp^2}$  and $L\gg L_m\equiv1/\sqrt[4]{f_2}\bp$, which regard the boundary conditions for the spin-2 and matter sectors, respectively~\cite{Salvio:2019ewf}. The possible instability occurring when such condition is violated not only is compatible with cosmology, but would also explain why we live in a nearly homogeneous and isotropic universe~\cite{Salvio:2019ewf}: the instability can take place only in sufficiently inhomogeneous and anisotropic patches of the universe.
\item Sec.~\ref{Euclidean path integral and Dirac-Pauli variables} was devoted to the application of the DP quantization and consequent norm operators to quadratic gravity to obtain a unitary theory (all probabilities are non negative and they sum up to one at any time, Eqs.~(\ref{ProbNorm}) and~(\ref{GenUnit})). \\ 
The Euclidean path integral for transition amplitudes and T-product matrix elements was constructed for a generic theory containing DP canonical variables in Sec.~\ref{EPTTp}. It was observed that the consistency of the Euclidean path integral puts strong constraints on the possible quantizations and we found one that works for quadratic gravity. As a byproduct, we also quantized in a similar manner a scalar-field four-derivative interacting model, whose Lagrangian is defined in~(\ref{ScalLag}). The quantization is summarized in Table~\ref{DPvsOrd} with all canonical coordinates on the right of the corresponding momenta in the quantum Hamiltonian $H$. The canonical coordinates that are invariant under time-reversal $T$ are quantized in the ordinary way, while those that are $T$-odd are quantized \`a la DP as proposed in~\cite{Salvio:2015gsi}. The probability density for the canonical coordinates is positive everywhere and its integral equals 1 at any time as required by unitarity, see~(\ref{unitaryrho}). An explicit expression for the measure, as well as the integrand, of the path integral was obtained in Sec.~\ref{EPTTp} and Appendix~\ref{Gaussian integral over gravitational momenta}. Like in ordinary quantum field theories, bare parameters are restricted by the convergence of the Euclidean path integral, but there remains a large region of available parameter space, contrary to GR (without the quadratic in curvature terms).  \\ 
As already mentioned, the description of the system in a generic coordinate system was provided  in Sec.~\ref{Transition to generic coordinate systems}. This is achieved in terms of a generic gauge-fixing function $f$ like in the FP approach to quantize ordinary gauge theories.
\\
In Sec.~\ref{EPTTpp} the Euclidean path integral for Green's functions is obtained. Whenever an ordinary two-derivative theory is recovered in the large-distance limit (which is the case both in quadratic gravity and in the discussed scalar-field model), this can be done by using the path integral for the matrix elements in~(\ref{PIEuclidean}). It was found that the large-distance spectrum of $H$ is real and bounded from below; also, the corresponding smallest eigenvalue $E_0$ is non negative if the bare cosmological constant $\lambda$ (in the case of quadratic gravity) or $V(0)$ (in the scalar-field model) is large enough. In the same section also the Lorentzian path  integral for Green's functions was obtained, which justifies previous expressions used in perturbative calculations around the flat spacetime.
\item The (quantum) effective action  $\Gamma$ for quadratic gravity in the specified quantization is defined in Sec.~\ref{The quadratic gravity effective action}. When used at tree level $\Gamma$ captures the full quantum effects and its field equations allow us to determine the VEV of the metric in the presence of a generic (possibly vanishing) physical energy-momentum tensor. 
\item In the classical limit, $\hbar\to0$, the field equations generated by the (quantum) effective action reduce to those of the classical action of quadratic gravity, as shown in Sec.~\ref{The classical limit}.  In the same section it was also explained how this theory, which features a large-distance spectrum of $H$ bounded from below, can have a classical Hamiltonian $H_c$ that is instead unbounded from below.
\item Finally, in Sec.~\ref{Quantum cosmology} the constructed non-perturbative and background-independent quantum quadratic gravity was applied to cosmology. 
\\
First, it was noted that the theory reduces (and thus justifies) previous tree-level calculations performed expanding around cosmological FLRW backgrounds. 
\\
Next, the consistent Euclidean path integral of quadratic gravity was applied to quantum cosmology. Such path integral, indeed, allows us to find an expression for the ground-state wave function of the universe and for all the large-distance spectrum. The wave functions (which are actually functionals of the metric and its time derivative) satisfy a sort of Wheeler-DeWitt equation. In our fully quantum construction the configuration that leads to the smallest action corresponds to peaks in the ground-state wave functional that are homogeneous and isotropic, while inhomogeneities and anisotropies  lead to  larger actions and are associated with excited states. These results support those of 
 Refs.~\cite{Lehners:2019ibe,Lehners:2023fud}, where the finite-action principle was proposed to explain homogeneity and isotropy of the universe within quadratic gravity in the classical limit.
Furthermore, it was found that large inhomogeneities and anisotropies necessarily correspond to large values of (the real part of) the eigenvalues of $H$. 
When the coefficient of the $W^2$ vanishes one recovers Starobinsky's model of inflation, which can, therefore, be studied  non-perturbatively with our construction. In the presence of the $W^2$
we are unable to show whether $H$ is diagonalizable with only real eigenvalues or not. However, in this final section we found that if that were not the case there would be no way to observe such unusual features: patches where their effects are sizable are not compatible with observers. It is only in the nearly homogeneous and isotropic  patches that inflation can occur and observations can eventually be made.  This mechanism is an explicit quantum realization of Linde's chaotic inflation~\cite{Linde:1983gd} and upgrades to the full quantum level the classical results of Ref.~\cite{Salvio:2019ewf}, which explain the near homogeneity and isotropy of our universe.
\end{itemize}

\subsubsection*{Acknowledgments} 
This work was partially supported by the Italian Ministry of University and Research (MUR) under the grant PNRR-M4C2-I1.1-PRIN 2022-PE2 Non-perturbative aspects of fundamental interactions, in the Standard Model and beyond F53D23001480006 funded by E.U. - NextGenerationEU and by the grant DyConn from the University of Rome Tor Vergata.

\appendix

\section{Curvature tensors in the  $3+1$ formalism}\label{Curvature tensors}

Here we give the non-vanishing independent components of the curvature tensors (the Riemann and the Ricci tensors) in the $3+1$ formalism and using the Gauss coordinates in~(\ref{GaussCo}):
\bea && R_{i0~j}^{~~0} = \dot K_{ij}+K_{il}K^l_{~j}, \quad  R_{0i~l}^{~~j} = D^jK_{il}-D_lK^j_{~i}, \\
&& R_{li~~j}^{~~m}= \, R^{\3~\,\,m}_{li~~j} +K_l^{~m} K_{ij}-K_i^{~m} K_{lj}, \\
&& R_{00} = g^{ij}(\dot K_{ij}+K_{il}K^l_{~j}), \quad R_{0i} = D_i K-D^jK_{ji}, \\ 
&& R_{ij} = \, R^\3_{ij} -\dot K_{ij}+ KK_{ij}-2K_{il} K^l_{~j},  \eea
where a zero denotes the time component. All the other components of the curvature tensors can be obtained by using the well-known properties under exchanges of their indices.  
These formul\ae~allow us to calculate the invariants in Eqs.~(\ref{R31}),~(\ref{Rmunu31}),~(\ref{Rmunurhosigma31}).

\section{Integral over gravitational momenta}\label{Gaussian integral over gravitational momenta}
In this appendix the Gaussian integral~(\ref{intetaij}) is computed. 
First, note that the ``matrix" appearing there  can be rewritten as 
\be\Gamma_{pqij} = g_{pm}g_{qn}~ \Gamma^{mn}_{\quad\, ij} \ee
where
\be \Gamma^{mn}_{\quad\, ij}  =  \frac1{\alpha} \frac{\delta^m_{i}\delta^n_{j}+\delta^m_{j}\delta^n_{i}}2 - \frac{4\beta-\alpha/3}{12\beta\alpha} g^{mn}g_{ij}.\ee
Now, let us compute the determinant of this ``matrix", which allows us to calculate, as well known, the corresponding Gaussian integral. We will first compute the determinant $\det\{\Gamma^{mn}_{\quad\, ij}\}$ of the ``matrix" with elements $\Gamma^{mn}_{\quad\, ij}$ and then that of $g_{pm}g_{qn}$, which will be denoted $\det\{g_{pm}g_{qn}\}$; finally, we will make the product of the two.

Note that, at each spacetime point,  $\Gamma^{mn}_{\quad\, ij}$ can be viewed as an operator from the (six-dimensional) vector space of symmetric $3\times 3$ matrices, $S^{ij}$, in itself. We can always decompose 
\be S^{ij} = S^{ij}_t +\frac{1}{3} S g^{ij}, \ee 
where $S\equiv g_{ij} S^{ij}$ is the trace and $S^{ij}_t \equiv S^{ij}-\frac{1}{3} S g^{ij}$ is the traceless part. The eigenvalue equation for the ``matrix" $\Gamma^{mn}_{\quad\, ij}$
\be \Gamma^{mn}_{\quad\, ij} S^{ij} = \lambda S^{mn}  \ee
 then reads
 \be \frac{S^{mn}_t}{\alpha}+ \frac{S g^{mn}}{36\beta}  =  \lambda (S^{mn}_t +\frac{1}{3} S g^{mn}).\ee 
 The space of traceless symmetric tensors $S^{ij}_t$ is five dimensional, while that of the trace $S$ is one dimensional so  the determinant of the \	``matrix" with elements $\Gamma^{mn}_{\quad\, ij}$ is 
 \be \det\{\Gamma^{mn}_{\quad\, ij}\} = \frac1{12\beta\alpha^5}, \label{1stdet}\ee 
 which is spacetime independent.
 
 The determinant of the other matrix factor, $g_{pm}g_{qn}$, can be computed by considering, at each spacetime point, a complete set of three eigenvectors $V_k$, with corresponding eigenvalues $\lambda_k$, of the metric:
 \be \sum_m g_{im} V_k^m = \lambda_k V_k^i, \label{gV}\ee
 where in the right-hand side there is no sum over $k$.
 Indeed, $(V_k^mV_l^n+V_l^mV_k^n)/2$ forms a set of six linearly-independent
  symmetric $3\times 3$ matrices, which are eigenvectors of the ``matrix" in question:
 \be \sum_{mn}g_{pm}g_{qn} \frac{V_k^mV_l^n+V_l^mV_k^n}2 = \lambda_k\lambda_l\frac{V_k^pV_l^q+V_l^pV_k^q}2. \label{ggV}\ee 
So, 
 \be \det \left\{ g_{pm}g_{qn} \right\}  = \prod_{k\geq l} \lambda_k\lambda_l =  g^4. \label{detgV}\ee
Combining this result with~(\ref{1stdet}) leads to
\be  \det \left\{\Gamma_{pqij}\right\} = \frac{g^4}{12\beta\alpha^5}.  \ee

By applying the Gaussian integration formula we then find
\be C' = \prod_{x_E'} \left[\left(\frac{\Delta V_3 \sqrt{g}}{\pi\hbar\Delta\tau}\right)^3 \frac{2\sqrt{3\beta\alpha^5}}{g^2}\right].  \label{Cpres} \ee
 \vspace{1cm}

\footnotesize
\begin{multicols}{2}

\end{multicols}


\begin{thebibliography}{}
\small{
 
 %\cite{LIGOScientific:2016aoc}
\bibitem{LIGOScientific:2016aoc}
B.~P.~Abbott \textit{et al.} [LIGO Scientific and Virgo],
``Observation of Gravitational Waves from a Binary Black Hole Merger,''
Phys. Rev. Lett. \textbf{116} (2016) no.6, 061102
doi:10.1103/PhysRevLett.116.061102
[\hhref{1602.03837}].

%\cite{EventHorizonTelescope:2019dse}
\bibitem{EventHorizonTelescope:2019dse}
K.~Akiyama \textit{et al.} [Event Horizon Telescope],
``First M87 Event Horizon Telescope Results. I. The Shadow of the Supermassive Black Hole,''
Astrophys. J. Lett. \textbf{875} (2019), L1
doi:10.3847/2041-8213/ab0ec7
[\hhref{1906.11238}].

%\cite{EventHorizonTelescope:2019ggy}
\bibitem{EventHorizonTelescope:2019ggy}
K.~Akiyama \textit{et al.} [Event Horizon Telescope],
``First M87 Event Horizon Telescope Results. VI. The Shadow and Mass of the Central Black Hole,''
Astrophys. J. Lett. \textbf{875} (2019) no.1, L6
doi:10.3847/2041-8213/ab1141
[\hhref{1906.11243}].

%\cite{EventHorizonTelescope:2022xnr}
\bibitem{EventHorizonTelescope:2022xnr}
K.~Akiyama \textit{et al.} [Event Horizon Telescope],
``First Sagittarius A* Event Horizon Telescope Results. I. The Shadow of the Supermassive Black Hole in the Center of the Milky Way,''
Astrophys. J. Lett. \textbf{930} (2022) no.2, L12
doi:10.3847/2041-8213/ac6674

%\cite{EventHorizonTelescope:2022wok}
\bibitem{EventHorizonTelescope:2022wok}
K.~Akiyama \textit{et al.} [Event Horizon Telescope],
``First Sagittarius A* Event Horizon Telescope Results. III. Imaging of the Galactic Center Supermassive Black Hole,''
Astrophys. J. Lett. \textbf{930} (2022) no.2, L14
doi:10.3847/2041-8213/ac6429

%\cite{EventHorizonTelescope:2022xqj}
\bibitem{EventHorizonTelescope:2022xqj}
K.~Akiyama \textit{et al.} [Event Horizon Telescope],
``First Sagittarius A* Event Horizon Telescope Results. VI. Testing the Black Hole Metric,''
Astrophys. J. Lett. \textbf{930} (2022) no.2, L17
doi:10.3847/2041-8213/ac6756
%28 citations counted in INSPIRE as of 16 Jun 2022

%\cite{Goroff:1985sz}
\bibitem{Goroff:1985sz}
M.~H.~Goroff and A.~Sagnotti,
``Quantum Gravity at Two Loops,''
Phys. Lett. B \textbf{160} (1985), 81-86
doi:10.1016/0370-2693(85)91470-4


%\cite{Goroff:1985th}
\bibitem{Goroff:1985th}
M.~H.~Goroff and A.~Sagnotti,
``The Ultraviolet Behavior of Einstein Gravity,''
Nucl. Phys. B \textbf{266} (1986), 709-736
doi:10.1016/0550-3213(86)90193-8

%\cite{Salvio:2018crh}
\bibitem{Salvio:2018crh}
A.~Salvio,
``Quadratic Gravity,''
Front. in Phys. \textbf{6} (2018), 77
doi:10.3389/fphy.2018.00077
[\hhref{1804.09944}].

%\cite{Donoghue:2021cza}
\bibitem{Donoghue:2021cza}
J.~F.~Donoghue and G.~Menezes,
``On quadratic gravity,''
Nuovo Cim. C \textbf{45}, no.2, 26 (2022)
doi:10.1393/ncc/i2022-22026-7
[\hhref{2112.01974}].


%\cite{Weinberg:1974tw}
\bibitem{Weinberg:1974tw}
 S.~Weinberg,
``Problems in Gauge Field Theories.'' In the proceedings of the XVII International Conference on High Energy Physics, editor J.~R.~Smith (Rutherford Laboratory, Chilton, Didcot, Oxfordshire), III-59.
  
  %\cite{Deser:1975nv}
\bibitem{Deser:1975nv}
  S.~Deser,
  ``The State of Quantum Gravity,''
  Conf.\ Proc.\ C {\bf 750926} (1975) 229. In the proceedings of the conference on Gauge Theories and Modern Field Theory, editors R.~Arnowitt and P.~Nath (MIT press, Cambridge, Massachusetts).
  %%CITATION = CONFP,C750926,229;%%
  
  %\cite{Stelle:1976gc}
\bibitem{Stelle:1976gc}
 K.~S.~Stelle,
``Renormalization of Higher Derivative Quantum Gravity,''
Phys. Rev. D \textbf{16} (1977), 953-969
doi:10.1103/PhysRevD.16.953

%\cite{Barvinsky:2017zlx}
\bibitem{Barvinsky:2017zlx}
A.~O.~Barvinsky, D.~Blas, M.~Herrero-Valea, S.~M.~Sibiryakov and C.~F.~Steinwachs,
``Renormalization of gauge theories in the background-field approach,''
JHEP \textbf{07}, 035 (2018)
doi:10.1007/JHEP07(2018)035
[\hhref{1705.03480}].

  \bibitem{Gibbons:1994cg}
  G.~W.~Gibbons and S.~W.~Hawking,
  ``Euclidean quantum gravity,'' Singapore, Singapore: World Scientific (1993) 586 p
  
  %\cite{Menotti:1989ms}
\bibitem{Menotti:1989ms}
P.~Menotti,
``Nonperturbative quantum gravity,''
Nucl. Phys. B Proc. Suppl. \textbf{17}, 29-38 (1990)
doi:10.1016/0920-5632(90)90218-J

  %\cite{Salvio:2015gsi}
\bibitem{Salvio:2015gsi}
A.~Salvio and A.~Strumia,
``Quantum mechanics of 4-derivative theories,''
Eur. Phys. J. C \textbf{76} (2016) no.4, 227
doi:10.1140/epjc/s10052-016-4079-8
[\hhref{1512.01237}].

 \bibitem{Pais}
A.~Pais and G.~E.~Uhlenbeck,  ``On Field theories with nonlocalized action'',  Phys.\ Rev.\  {79} (1950) 145.

%\cite{Pauli}
\bibitem{Pauli}  W.~Pauli, ``On Dirac's New Method of Field Quantization", Rev.\ Mod.\ Phys.\ 15 (1943) 175
doi:10.1103/RevModPhys.15.175 
      
   %\cite{Dirac}
\bibitem{Dirac}    P.~A.~M.~Dirac, ``The physical interpretation of quantum mechanics," 
	Proc.\ R.\ Soc.\ Lond.\ A 180, 1 (1942)
	doi:10.1098/rspa.1942.0023
	
	%\cite{Buchbinder-Lyahovich:1987}
\bibitem{Buchbinder-Lyahovich:1987}
Canonical quantisation and local measure of R2 gravity
I.~L.~Buchbinder and S.~L.~Lyahovich,
``Canonical quantisation and local measure of $R^2$ gravity"
 Class. Quantum Grav. \textbf{4} (1987)  1487
doi:10.1088/0264-9381/4/6/008

%\cite{Buchbinder:1991ne}
\bibitem{Buchbinder:1991ne}
I.~L.~Buchbinder, I.~Y.~Karataeva and S.~L.~Lyakhovich,
``Multidimensional R**2 gravity: The Structure of constraints and canonical quantization,''
Class. Quant. Grav. \textbf{8}, 1113-1125 (1991)
doi:10.1088/0264-9381/8/6/010

%\cite{Buchbinder:1992rb}
\bibitem{Buchbinder:1992rb}
  I.~L.~Buchbinder, S.~D.~Odintsov and I.~L.~Shapiro,
  ``Effective action in quantum gravity,''
  Bristol, UK: IOP (1992) 413 p
  
  %\cite{Strumia:2017dvt}
\bibitem{Strumia:2017dvt}
A.~Strumia,
``Interpretation of quantum mechanics with indefinite norm,''
MDPI Physics \textbf{1} (2019) no.1, 17-32
doi:10.3390/physics1010003
[\hhref{1709.04925}].

	
%\cite{Salvio:2019wcp}
\bibitem{Salvio:2019wcp}
A.~Salvio,
``Quasi-Conformal Models and the Early Universe,''
Eur. Phys. J. C \textbf{79} (2019) no.9, 750
doi:10.1140/epjc/s10052-019-7267-5
[\hhref{1907.00983}].

%\cite{Salvio:2020axm}
\bibitem{Salvio:2020axm}
A.~Salvio,
``Dimensional Transmutation in Gravity and Cosmology,''
Int. J. Mod. Phys. A \textbf{36} (2021) no.08n09, 2130006
doi:10.1142/S0217751X21300064
[\hhref{2012.11608}].

%\cite{Bender:2007wu}
\bibitem{Bender:2007wu}
C.~M.~Bender and P.~D.~Mannheim,
``No-ghost theorem for the fourth-order derivative Pais-Uhlenbeck oscillator model,''
Phys. Rev. Lett. \textbf{100}, 110402 (2008)
doi:10.1103/PhysRevLett.100.110402
[\hhref{0706.0207}].

%\cite{Bender:2008gh}
\bibitem{Bender:2008gh}
C.~M.~Bender and P.~D.~Mannheim,
``Exactly solvable PT-symmetric Hamiltonian having no Hermitian counterpart,''
Phys. Rev. D \textbf{78}, 025022 (2008)
doi:10.1103/PhysRevD.78.025022
[\hhref{0804.4190}].

%\cite{Bender:2023cem}
\bibitem{Bender:2023cem}
C.~M.~Bender and D.~W.~Hook,
``PT-symmetric quantum mechanics,''
[\hhref{2312.17386}].

%\cite{Salvio:2017xul}
\bibitem{Salvio:2017xul}
A.~Salvio,
``Inflationary Perturbations in No-Scale Theories,''
Eur. Phys. J. C \textbf{77} (2017) no.4, 267
doi:10.1140/epjc/s10052-017-4825-6
[\hhref{1703.08012}].



    \bibitem{ostro}
M. Ostrogradsky, ``Memoires sur les \'equations diff\'erentielles relatives au probl\`eme des isop\'erim\`etres," Mem. Ac. St. Petersbourg VI (1850) 385. \href{https://babel.hathitrust.org/cgi/pt?id=mdp.39015038710128;view=1up;seq=405}{Pdf available online}.

%\cite{Salvio:2019ewf}
\bibitem{Salvio:2019ewf}
A.~Salvio,
``Metastability in Quadratic Gravity,''
Phys. Rev. D \textbf{99} (2019) no.10, 103507
doi:10.1103/PhysRevD.99.103507
[\hhref{1902.09557}].


%\cite{Gross:2020tph}
\bibitem{Gross:2020tph}
C.~Gross, A.~Strumia, D.~Teresi and M.~Zirilli,
``Is negative kinetic energy metastable?,''
Phys. Rev. D \textbf{103} (2021) no.11, 115025
doi:10.1103/PhysRevD.103.115025
[\hhref{2007.05541}].

%\cite{Held:2021pht}
\bibitem{Held:2021pht}
A.~Held and H.~Lim,
``Nonlinear dynamics of quadratic gravity in spherical symmetry,''
Phys. Rev. D \textbf{104} (2021) no.8, 084075
doi:10.1103/PhysRevD.104.084075
[\hhref{2104.04010}].

%\cite{Held:2023aap}
\bibitem{Held:2023aap}
A.~Held and H.~Lim,
``Nonlinear evolution of quadratic gravity in 3+1 dimensions,''
Phys. Rev. D \textbf{108}, no.10, 104025 (2023)
doi:10.1103/PhysRevD.108.104025
[\hhref{2306.04725}].

%\cite{Deffayet:2021nnt}
\bibitem{Deffayet:2021nnt}
C.~Deffayet, S.~Mukohyama and A.~Vikman,
``Ghosts without Runaway Instabilities,''
Phys. Rev. Lett. \textbf{128}, no.4, 041301 (2022)
doi:10.1103/PhysRevLett.128.041301
[\hhref{2108.06294}].

%\cite{Deffayet:2023wdg}
\bibitem{Deffayet:2023wdg}
C.~Deffayet, A.~Held, S.~Mukohyama and A.~Vikman,
``Global and local stability for ghosts coupled to positive energy degrees of freedom,''
JCAP \textbf{11}, 031 (2023)
doi:10.1088/1475-7516/2023/11/031
[\hhref{2305.09631}].

%\cite{Lee:1969fy}
\bibitem{Lee:1969fy}
T.~D.~Lee and G.~C.~Wick,
``Negative Metric and the Unitarity of the S Matrix,''
Nucl. Phys. B \textbf{9}, 209-243 (1969)
doi:10.1016/0550-3213(69)90098-4

%\cite{Salvio:2018kwh}
\bibitem{Salvio:2018kwh}
A.~Salvio, A.~Strumia and H.~Veerm\"ae,
``New infra-red enhancements in 4-derivative gravity,''
Eur. Phys. J. C \textbf{78} (2018) no.10, 842
doi:10.1140/epjc/s10052-018-6311-1
[\hhref{1808.07883}].


%\cite{Donoghue:2019fcb}
\bibitem{Donoghue:2019fcb}
J.~F.~Donoghue and G.~Menezes,
``Unitarity, stability and loops of unstable ghosts,''
Phys. Rev. D \textbf{100}, no.10, 105006 (2019)
doi:10.1103/PhysRevD.100.105006
[\hhref{1908.02416}].

%\cite{Holdom:2021hlo}
\bibitem{Holdom:2021hlo}
B.~Holdom,
``Ultra-Planckian scattering from a QFT for gravity,''
Phys. Rev. D \textbf{105}, no.4, 046008 (2022)
doi:10.1103/PhysRevD.105.046008
[\hhref{2107.01727}].

%\cite{Anselmi:2017ygm}
\bibitem{Anselmi:2017ygm}
  D.~Anselmi,
  ``On the quantum field theory of the gravitational interactions,''
  JHEP {\bf 1706} (2017) 086 
  [\hhref{1704.07728}].


%\cite{Hartle:1983ai}
\bibitem{Hartle:1983ai}
J.~B.~Hartle and S.~W.~Hawking,
``Wave Function of the Universe,''
Phys. Rev. D \textbf{28} (1983), 2960-2975
doi:10.1103/PhysRevD.28.2960

 %\cite{DeWitt:1967yk}
\bibitem{DeWitt:1967yk}
B.~S.~DeWitt,
``Quantum Theory of Gravity. 1. The Canonical Theory,''
Phys. Rev. \textbf{160} (1967), 1113-1148
doi:10.1103/PhysRev.160.1113

%\cite{Wheeler}
\bibitem{Wheeler}
 J. A. Wheeler, in Battelle Rencontres, edited by C. DeWitt and J. A. Wheeler (Benjamin, New York, 1968).

%\cite{Starobinsky:1980te}
\bibitem{Starobinsky:1980te}
A.~A.~Starobinsky,
``A New Type of Isotropic Cosmological Models Without Singularity,''
Phys. Lett. B \textbf{91}, 99-102 (1980)
doi:10.1016/0370-2693(80)90670-X

%\cite{Ade:2015lrj}
\bibitem{Ade:2015lrj}
  P.~A.~R.~Ade {\it et al.} [Planck Collaboration],
  ``Planck 2015 results. XX. Constraints on inflation,''
  Astron.\ Astrophys.\  {\bf 594} (2016) A20 
  doi:10.1051/0004-6361/201525898
  [\hhref{1502.02114}].

\bibitem{Planck2018:inflation}
Y.~Akrami, et~al., {Planck 2018 results. X. Constraints on inflation}, Astron.
  Astrophys. 641 (2020) A10
  doi:10.1051/0004-6361/201833887
 [\hhref{arXiv:1807.06211}].
  
  %\cite{BICEP:2021xfz}
\bibitem{BICEP:2021xfz}
P.~A.~R.~Ade \textit{et al.} [BICEP and Keck],
``Improved Constraints on Primordial Gravitational Waves using Planck, WMAP, and BICEP/Keck Observations through the 2018 Observing Season,''
Phys. Rev. Lett. \textbf{127} (2021) no.15, 151301
doi:10.1103/PhysRevLett.127.151301
[\hhref{2110.00483}].

 \bibitem{WeinbergAS} S. Weinberg, in Understanding the Fundamental Constituents of Matter, ed. A. Zichichi (Plenum 	Press, New York, 1977). S. Weinberg,  in General  Relativity: An  Einstein Centenary Survey, 	edited by S. W. Hawking and W. Israel (Cambridge University Press, 1980) pp. 790-831.
 
 %\cite{Benedetti:2009rx}
\bibitem{Benedetti:2009rx}
D.~Benedetti, P.~F.~Machado and F.~Saueressig,
``Asymptotic safety in higher-derivative gravity,''
Mod. Phys. Lett. A \textbf{24}, 2233-2241 (2009)
doi:10.1142/S0217732309031521
[\hhref{0901.2984}].

%\cite{Falls:2020qhj}
\bibitem{Falls:2020qhj}
K.~Falls, N.~Ohta and R.~Percacci,
``Towards the determination of the dimension of the critical surface in asymptotically safe gravity,''
Phys. Lett. B \textbf{810}, 135773 (2020)
doi:10.1016/j.physletb.2020.135773
[\hhref{2004.04126}].

 %\cite{Arnowitt:1959ah}
\bibitem{Arnowitt:1959ah}
R.~L.~Arnowitt, S.~Deser and C.~W.~Misner,
``Dynamical Structure and Definition of Energy in General Relativity,''
Phys. Rev. \textbf{116} (1959), 1322-1330
doi:10.1103/PhysRev.116.1322

%\cite{Arnowitt:1962hi}
\bibitem{Arnowitt:1962hi}
R.~L.~Arnowitt, S.~Deser and C.~W.~Misner,
``The Dynamics of general relativity,''
Gen. Rel. Grav. \textbf{40} (2008), 1997-2027
doi:10.1007/s10714-008-0661-1
[\hhref{gr-qc/0405109}].

%\cite{Kluson:2013hza}
\bibitem{Kluson:2013hza}
  J.~Kluson, M.~Oksanen and A.~Tureanu,
  ``Hamiltonian analysis of curvature-squared gravity with or without conformal invariance,''
  Phys.\ Rev.\ D {\bf 89} (2014) no.6,  064043
  [\hhref{1311.4141}].
  
  %\cite{Weinberg:1996kr}
\bibitem{Weinberg:1996kr}
S.~Weinberg,
``The quantum theory of fields. Vol. 2: Modern applications,''
Cambridge University Press, 2013,
ISBN 978-1-139-63247-8, 978-0-521-67054-8, 978-0-521-55002-4
doi:10.1017/CBO9781139644174

%\cite{Bona}
\bibitem{Bona}
C.~Bona, C.~Palenzuela-Luque and C.~Bona-Casas, ``Elements of Numerical Relativity and Relativistic Hydrodynamics." (Springer, Dordrecht Heidelberg, London, New York, 2009)

%\cite{Bender:2009mq}
\bibitem{Bender:2009mq}
C.~M.~Bender and P.~D.~Mannheim,
``PT symmetry and necessary and sufficient conditions for the reality of energy eigenvalues,''
Phys. Lett. A \textbf{374}, 1616-1620 (2010)
doi:10.1016/j.physleta.2010.02.032
[\hhref{0902.1365}].



%\cite{Salvio:2016mvj}
\bibitem{Salvio:2016mvj}
A.~Salvio, A.~Strumia, N.~Tetradis and A.~Urbano,
``On gravitational and thermal corrections to vacuum decay,''
JHEP \textbf{09}, 054 (2016)
doi:10.1007/JHEP09(2016)054
[\hhref{1608.02555}].

  
%\cite{Salvio:2014soa}
\bibitem{Salvio:2014soa}
A.~Salvio and A.~Strumia,
``Agravity,''
JHEP \textbf{06} (2014), 080
doi:10.1007/JHEP06(2014)080
[\hhref{1403.4226}].

%\cite{Salvio:2017qkx}
\bibitem{Salvio:2017qkx}
A.~Salvio and A.~Strumia,
``Agravity up to infinite energy,''
Eur. Phys. J. C \textbf{78} (2018) no.2, 124
doi:10.1140/epjc/s10052-018-5588-4
[\hhref{1705.03896}].

%\cite{Avramidi:1985ki}
\bibitem{Avramidi:1985ki}
I.~G.~Avramidi and A.~O.~Barvinsky,
``Asymptotic freedom in higher derivative quantum gravity,''
Phys. Lett. B \textbf{159}, 269-274 (1985)
doi:10.1016/0370-2693(85)90248-5

%\cite{Buccio:2024hys}
\bibitem{Buccio:2024hys}
D.~Buccio, J.~F.~Donoghue, G.~Menezes and R.~Percacci,
``Physical running of couplings in quadratic gravity,''
[\hhref{2403.02397}].

  %\cite{Deruelle:2010kf}
\bibitem{Deruelle:2010kf}
N.~Deruelle, M.~Sasaki, Y.~Sendouda and A.~Youssef,
``Inflation with a Weyl term, or ghosts at work,''
JCAP \textbf{03}, 040 (2011)
doi:10.1088/1475-7516/2011/03/040
[\hhref{1012.5202}].

%\cite{Kannike:2015apa}
\bibitem{Kannike:2015apa}
K.~Kannike, G.~H\"utsi, L.~Pizza, A.~Racioppi, M.~Raidal, A.~Salvio and A.~Strumia,
``Dynamically Induced Planck Scale and Inflation,''
JHEP \textbf{05} (2015), 065
doi:10.1007/JHEP05(2015)065
[\hhref{1502.01334}].



%\cite{Lu:2015cqa}
\bibitem{Lu:2015cqa}
H.~Lu, A.~Perkins, C.~N.~Pope and K.~S.~Stelle,
``Black Holes in Higher-Derivative Gravity,''
Phys. Rev. Lett. \textbf{114}, no.17, 171601 (2015)
doi:10.1103/PhysRevLett.114.171601
[\hhref{1502.01028}].

%\cite{Lu:2015psa}
\bibitem{Lu:2015psa}
H.~L\"u, A.~Perkins, C.~N.~Pope and K.~S.~Stelle,
``Spherically Symmetric Solutions in Higher-Derivative Gravity,''
Phys. Rev. D \textbf{92}, no.12, 124019 (2015)
doi:10.1103/PhysRevD.92.124019
[\hhref{1508.00010}].

%\cite{Lu:2017kzi}
\bibitem{Lu:2017kzi}
H.~L\"u, A.~Perkins, C.~N.~Pope and K.~S.~Stelle,
``Lichnerowicz Modes and Black Hole Families in Ricci Quadratic Gravity,''
Phys. Rev. D \textbf{96}, no.4, 046006 (2017)
doi:10.1103/PhysRevD.96.046006
[\hhref{1704.05493}].

%\cite{dePaulaNetto:2023vtg}
\bibitem{dePaulaNetto:2023vtg}
T.~de Paula Netto, B.~L.~Giacchini, N.~Burzill\`a and L.~Modesto,
``Regular black holes from higher-derivative and nonlocal gravity: The smeared delta source approximation,''
[\hhref{2308.12251}].

%\cite{Giudice:2014tma}
\bibitem{Giudice:2014tma}
G.~F.~Giudice, G.~Isidori, A.~Salvio and A.~Strumia,
``Softened Gravity and the Extension of the Standard Model up to Infinite Energy,''
JHEP \textbf{02}, 137 (2015)
doi:10.1007/JHEP02(2015)137
[\hhref{1412.2769}].

%\cite{Salvio:2016vxi}
\bibitem{Salvio:2016vxi}
A.~Salvio,
``Solving the Standard Model Problems in Softened Gravity,''
Phys. Rev. D \textbf{94}, no.9, 096007 (2016)
doi:10.1103/PhysRevD.94.096007
[\hhref{1608.01194}].

%\cite{Holdom:2016nek}
\bibitem{Holdom:2016nek}
B.~Holdom and J.~Ren,
``Not quite a black hole,''
Phys. Rev. D \textbf{95}, no.8, 084034 (2017)
doi:10.1103/PhysRevD.95.084034
[\hhref{1612.04889}].

%\cite{Holdom:2019bdv}
\bibitem{Holdom:2019bdv}
B.~Holdom,
``Not quite black holes at LIGO,''
Phys. Rev. D \textbf{101}, no.6, 064063 (2020)
doi:10.1103/PhysRevD.101.064063
[\hhref{1909.11801}].

%\cite{Salvio:2019llz}
\bibitem{Salvio:2019llz}
A.~Salvio and H.~Veerm\"ae,
``Horizonless ultracompact objects and dark matter in quadratic gravity,''
JCAP \textbf{02}, 018 (2020)
doi:10.1088/1475-7516/2020/02/018
[\hhref{1912.13333}].

%\cite{Aydemir:2020xfd}
\bibitem{Aydemir:2020xfd}
U.~Aydemir, B.~Holdom and J.~Ren,
``Not quite black holes as dark matter,''
Phys. Rev. D \textbf{102}, no.2, 024058 (2020)
doi:10.1103/PhysRevD.102.024058
[\hhref{2003.10682}].

%\cite{Donoghue:2021eto}
\bibitem{Donoghue:2021eto}
J.~F.~Donoghue and G.~Menezes,
``Ostrogradsky instability can be overcome by quantum physics,''
Phys. Rev. D \textbf{104} (2021) no.4, 045010
doi:10.1103/PhysRevD.104.045010
[\hhref{2105.00898}].


  
%\cite{Anselmi:2020lpp}
\bibitem{Anselmi:2020lpp}
D.~Anselmi, E.~Bianchi and M.~Piva,
``Predictions of quantum gravity in inflationary cosmology: effects of the Weyl-squared term,''
JHEP \textbf{07} (2020), 211
[\hhref{2005.10293}].

%\cite{Salvio:2022mld}
\bibitem{Salvio:2022mld}
A.~Salvio,
``BICEP/Keck data and quadratic gravity,''
JCAP \textbf{09} (2022), 027
doi:10.1088/1475-7516/2022/09/027
[\hhref{2202.00684}].

%\cite{Dondarini:2023atl}
\bibitem{Dondarini:2023atl}
A.~Dondarini,
``Causality bounds in quadratic inflation from purely virtual particles,''
Phys. Rev. D \textbf{108} (2023) no.8, 083526
doi:10.1103/PhysRevD.108.083526
[\hhref{2306.04687}].


%\cite{Hawking:1981gb}
\bibitem{Hawking:1981gb}
S.~W.~Hawking,
``The Boundary Conditions of the Universe,''
Pontif. Acad. Sci. Scr. Varia \textbf{48}, 563-574 (1982)
PRINT-82-0179 (CAMBRIDGE).

%\cite{Lehners:2023yrj}
\bibitem{Lehners:2023yrj}
J.~L.~Lehners,
``Review of the no-boundary wave function,''
Phys. Rept. \textbf{1022}, 1-82 (2023)
doi:10.1016/j.physrep.2023.06.002
[\hhref{2303.08802}].

%\cite{Maldacena:2024uhs}
\bibitem{Maldacena:2024uhs}
J.~Maldacena,
``Comments on the no boundary wavefunction and slow roll inflation,''
[\hhref{2403.10510}].


%\cite{Lehners:2019ibe}
\bibitem{Lehners:2019ibe}
J.~L.~Lehners and K.~S.~Stelle,
``A Safe Beginning for the Universe?,''
Phys. Rev. D \textbf{100} (2019) no.8, 083540
doi:10.1103/PhysRevD.100.083540
[\hhref{1909.01169}].

%\cite{Lehners:2023fud}
\bibitem{Lehners:2023fud}
J.~L.~Lehners and K.~S.~Stelle,
``Higher-Order Gravity, Finite Action, and a Safe Beginning for the Universe,''
[\hhref{2312.14048}].

%\cite{Edery:2023hxl}
\bibitem{Edery:2023hxl}
A.~Edery,
``Enlarging the symmetry of pure $R^2$ gravity, BRST invariance and its spontaneous breaking,''
[\hhref{2301.08638}].

%\cite{Hell:2023mph}
\bibitem{Hell:2023mph}
A.~Hell, D.~Lust and G.~Zoupanos,
``On the degrees of freedom of R$^{2}$ gravity in flat spacetime,''
JHEP \textbf{02}, 039 (2024)
doi:10.1007/JHEP02(2024)039
[\hhref{2311.08216}].

%\cite{Maldacena:2011mk}
\bibitem{Maldacena:2011mk}
J.~Maldacena,
``Einstein Gravity from Conformal Gravity,''
[\hhref{1105.5632}].

%\cite{Hinterbichler:2015soa}
\bibitem{Hinterbichler:2015soa}
K.~Hinterbichler and M.~Saravani,
``St\"uckelberg approach to quadratic curvature gravity and its decoupling limits,''
Phys. Rev. D \textbf{93}, 065006 (2016)
doi:10.1103/PhysRevD.93.065006
[\hhref{1508.02401}].

%\cite{Hell:2023rbf}
\bibitem{Hell:2023rbf}
A.~Hell, D.~Lust and G.~Zoupanos,
``On the ghost problem of conformal gravity,''
JHEP \textbf{08}, 168 (2023)
doi:10.1007/JHEP08(2023)168
[\hhref{2306.13714}].

%\cite{Buoninfante:2023ryt}
\bibitem{Buoninfante:2023ryt}
L.~Buoninfante,
``Massless and partially massless limits in Quadratic Gravity,''
JHEP \textbf{12}, 111 (2023)
doi:10.1007/JHEP12(2023)111
[\hhref{2308.11324}].

\bibitem{Linde:1983gd}
A.~D.~Linde,
``Chaotic Inflation,''
Phys. Lett. B \textbf{129} (1983), 177-181
doi:10.1016/0370-2693(83)90837-7



  
  }
  

\end{thebibliography}
\end{document}